\documentclass[10pt,preprint]{aastex}
\usepackage{float}

\shorttitle{Radio Background at long wavelengths}
\shortauthors{Patra et al.}

\begin{document}


\title{SARAS measurement of the Radio Background at long wavelengths}


\author{Nipanjana Patra\altaffilmark{1,2} }
\email{Contact author email: nipanjana@rri.res.in}
\author{Ravi  Subrahmanyan }
\author{Shiv Sethi }
\author{N. Udaya  Shankar }
\author{A. Raghunathan }
\affil{Raman Research Institute, CV Raman Avenue, Sadashivanagar, Bangalore 560080, India}

\altaffiltext{1}{Present address: University of California, Berkeley, nipanjana@berkeley.edu}
\altaffiltext{2}{Joint Astronomy Program, Indian Institute of Science, Bangalore, India}


\begin{abstract}
SARAS is a correlation spectrometer connected to a frequency independent antenna that is purpose-designed for precision measurements of the radio background at long wavelengths.  The design, calibration and observing strategies admit solutions for the internal additive contributions to the radiometer response, and hence a separation of these contaminants from the antenna temperature.  We present here a wideband measurement of the radio sky spectrum by SARAS that provides an accurate measurement of the absolute brightness and spectral index between 110 and 175~MHz.  Accuracy in the measurement of absolute sky brightness is limited by systematic errors of magnitude 1.2\%; errors in calibration and in the joint estimation of sky and system model parameters are relatively smaller.  We use this wide-angle measurement of the sky brightness using the precision wide-band dipole antenna to provide an improved absolute calibration for the 150-MHz all-sky map of \citet{Landecker70}: subtracting an offset of 21.4~K and scaling by factor 1.05 will reduce the overall offset error to 8~K (from 50~K) and scale error to 0.8\% (from 5\%).  The SARAS measurement 
of the temperature spectral index is in the range $-2.3$--$-2.45$ in the 110-175~MHz band and indicates that the region towards the Galactic bulge has a relatively flatter index. \\
\end{abstract}


\keywords{Astronomical instrumentation, methods and techniques - Methods: observational - Background radiation - Radio continuum: ISM }

\section{Introduction}

The radio background at long wavelength consists of thermal and non-thermal emission from the different components of the Galaxy, radio emission from extragalactic radio sources and the cosmic microwave background (CMB).    Precision measurements of the absolute temperature and spectral distribution of the radio background are essential for the modeling of the Galactic components and constraining the extragalactic radio brightness, which would answer the question of whether there is any unaccounted uniform radio background (see \citet{Ravi13} and references therein).

Absolute calibration of the radio sky has usually been carried out using antenna elements that are amenable to precise characterization, which are connected to calibrated radiometers.  The brightness temperature of the radio background has been measured earlier at discrete frequencies using aerials that are scaled geometrically, and with wide-angle beams that average over different parts of the Galactic plane and off the plane.  

At long wavelengths, all sky images of the absolute brightness temperature distribution have been made at 150~MHz \citep{Landecker70} and at 408~MHz \citep{Haslam82}.  Partial coverage maps have been made at numerous other radio frequencies, a compilation is in \citet{deCosta08}.  Any modeling of the radio background based on these all sky maps as well as by combining the absolute brightness temperature measurements at discrete frequencies is limited by the absolute calibration errors; for example, the zero point error in the 408 MHz all sky map is $\pm3$~K and this increases to about $\pm40$~K in the 150~MHz map. The error in the absolute scales of these measurements varies from 1 to 10$\%$ with larger errors at lower frequencies.  The zero point uncertainty in the lower frequency measurement by \citet{Purton66} at 81.5~MHz is $\pm30$~K. Modeling of the radio background using these measurements are limited by the errors of calibration in individual measurements. 

Estimates of the spectral index of the radio background have mostly relied on interpolations between the maps available at various frequencies, with errors that are often dominated by the zero point offsets in the individual maps.  WMAP all sky measurement results in Galactic synchrotron emission spectral index between 408~MHz  to 23 GHz, which is flattest at the  galactic plane where the index $\alpha = -2.5$ (sky brightness $T \sim \nu^{\alpha}$, where $\nu$ is the frequency) and steepest towards the poles where $\alpha = -3.0$. \citep{Bennett03}.  The measurement by \cite{Costain60} at 38 and 178~MHz yields a spectral index of $-2.37\pm0.04$ for the whole of the northern  sky barring the central region of the galactic plane. The radio background spectrum measured by \cite{Turtle62} shows the spectral index to be ranging from $-2.0$ to $-2.9$  with the spectra along the direction of the galactic poles having spectral index slightly steeper than that towards the galactic anti-center. 

\begin{figure*}[ht]
\centering
\includegraphics[ width=12cm]{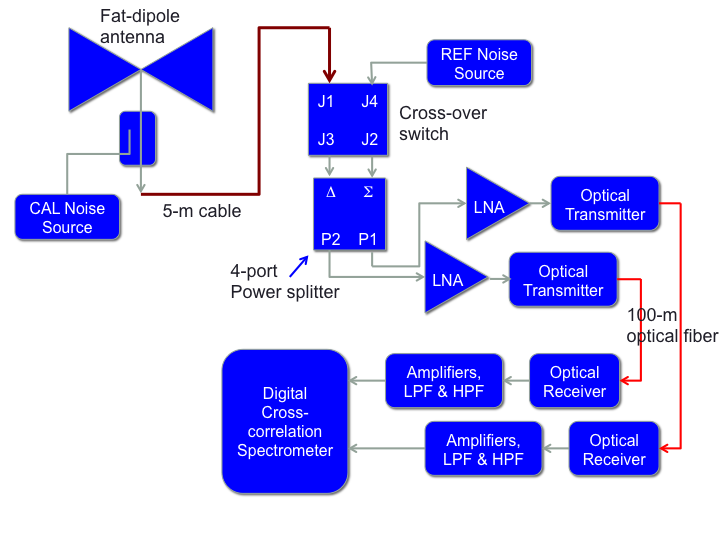}
\caption{The configuration of the SARAS spectrometer.}
\label{fig:design0}       
\end{figure*}

These measurements also show that the radio background spectrum has a curvature and the temperature spectral index is flatter at lower frequencies.  
\citet{Purton66} show that the radio background spectrum has a temperature spectral index of $-2.38\pm0.05$ between frequencies 13 to 100~MHz and increases to about $-2.9$ at 400~MHz. Combining measurements at frequencies between 10 and 38~MHz with the measurements of \cite{Purton66}, \cite{Andrew66}  determines a temperature spectral index of $-2.43\pm0.03$ between 10 to 178 MHz.  More accurate measurements of the spectral index are possible today with wide band antennas and receivers. Recently, \citet{RogBow2008} used a 4-point antenna and a radiometric spectrometer to estimate the mean temperature spectral index of the 100--200~MHz sky to be $\alpha = -2.5 \pm 0.1$. 

SARAS (Shaped Antenna Measurement of background RAdio Spectrum) is a ground based experiment designed to make precise measurements of the absolute spectrum of the radio background at long wavelengths \citep{Patra13} and over octave bandwidths. SARAS uses a wideband frequency-independent antenna, designed to operate in the 87.5--175~MHz band, and a correlation spectrometer with switching to cancel internal additive noise and provide means of calibration of spurious responses arising from internal reflections.  In this paper we describe SARAS measurements of the absolute spectrum of the radio background that have been made over the 110--170~MHz band. The sky was observed with SARAS on August 2 and 3, 2013 over the LST range 23~h to 1~h; i.e, when the meridian value of right ascension drifts from $23^{\rm h}$ to $01^{\rm h}$. In Section~2 we describe the system configuration and in Section~3 the measurement equations. Section~4 describes the method of absolute calibration. The hierarchical modeling of the data, the estimates of the sky temperature spectral index and the errors in the estimates are in Section~5.    

\section{SARAS system configuration} 

SARAS is a spectral radiometer that measures the difference between the antenna temperature and the noise temperature of a reference load (Fig.~\ref{fig:design0}). The antenna element is an octave-bandwidth frequency-independent fat-dipole of length $\approx 1.0$~m which correspond to the the half wavelength at 148~MHz \citep{Raghu}, and a switched calibration noise is added to the signal path via a $-20$~dB directional coupler that immediately follows the antenna.  A 5-m cable connects this combined signal to a cross-over switch, whose second input receives a switched reference noise power that serves as the reference load. The cross-over switch connects the antenna and the reference to the $\Delta$ and $\Sigma$ ports of a $180\degr$ power splitter alternately. 

The outputs of the splitter are fed to two identical receiver chains.  Signals along both paths are amplified by low-noise amplifiers (LNAs) and modulate optical carriers  for transmission via optic fibers to a base station 100~m away.  The demodulated electrical voltage signals are band limited and amplified. These signals are sampled at 175~MHz and discrete Fourier transforms are computed at 1024 frequency points across the 87.5--175~MHz band. The complex data in the corresponding frequency channels are then multiplied and integrated resulting in a complex cross-power spectrum over the observing band.  

\begin{figure}[ht]
\centering
\includegraphics[height=12cm, width=8cm]{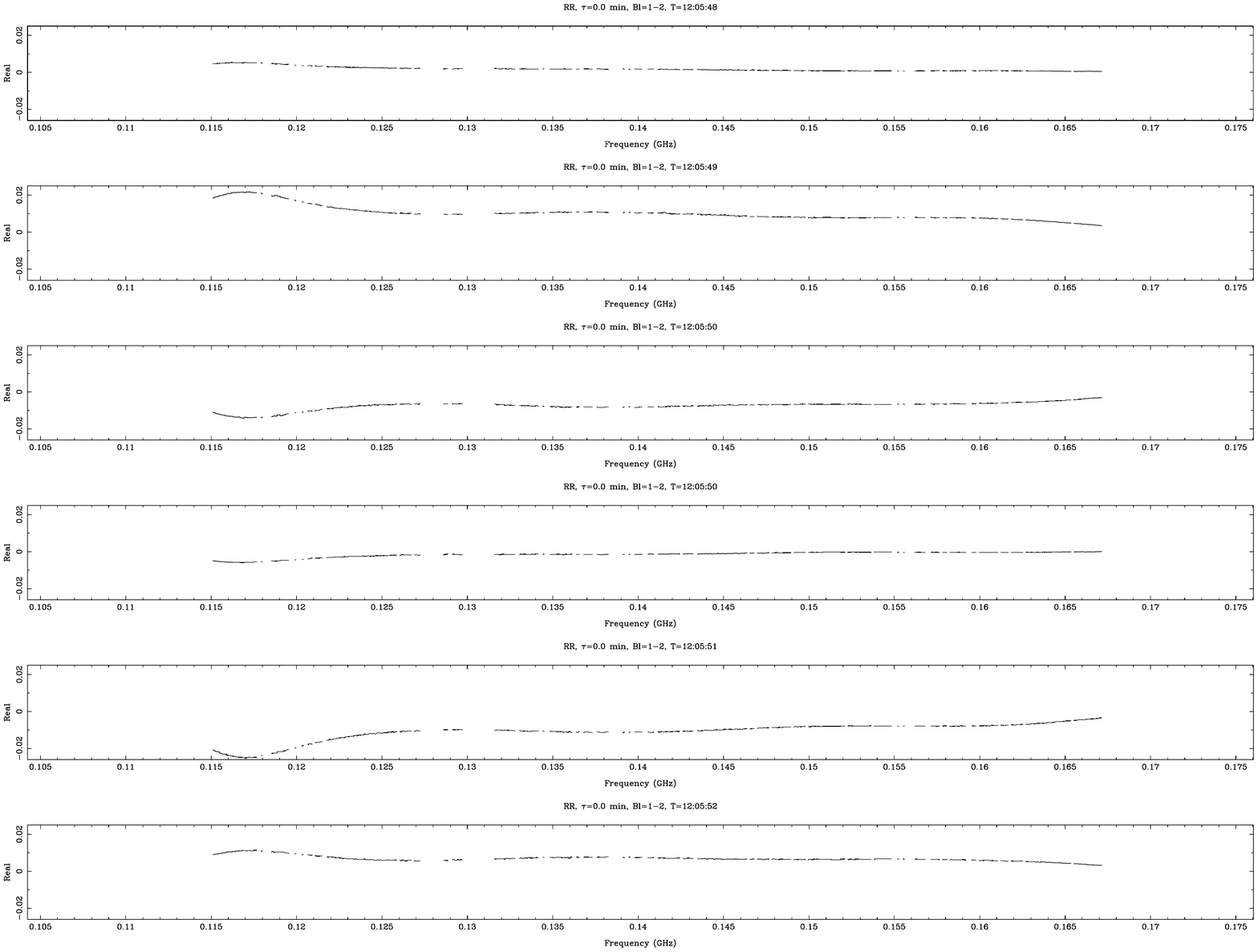}
\caption{Sample cross-power spectra recorded in a cycle of the observing sequence; only real components of cross spectra are displayed. The first row shows the spectra measured with the switch in position `0' and the second shows the records made with the switch in position `1'; in each row, the first is with both noise sources off, the second with calibration source alone on and the last with reference source alone on. (see Table~\ref{table:states})}
\label{raw_sample}       
\end{figure}

The cross power response to the antenna and the reference noise have opposite signs and hence the net cross power  is the difference between the antenna power and that from the reference.  The sign of this difference measurement flips in the two switch positions.  In each switch position three cross power spectra are recorded: first, when both noise sources are `off' and subsequently with the calibration noise source and reference noise source turned `on' one at a time.   Thus the system cycles through six different states as listed in Table~\ref{table:states} and in each state the power spectra corresponding to the two individual receiver chains as well as the complex cross-power spectrum are recorded. A set of example spectra is shown in Fig.~\ref{raw_sample}.  

\begin{table}[ht]
\caption{The SARAS observing cycle.  The six rows show the switch positions and on (1) and off (0) states of the calibrator and reference noise sources in the six states through which the system cycles during the observations.}
\label{table:states}
\centering
\begin{tabular}{lccc}
\noalign{\medskip}
\hline\noalign{\smallskip}
State Name & Switch state & CAL noise state & REF noise state  \\
\noalign{\smallskip}\hline\noalign{\smallskip}
OBS0 & 0 & 0 & 0 \\
CAL0 & 0 & 1 & 0 \\
REF0 & 0 & 0 & 1 \\
OBS1 & 1 & 0 & 0 \\
CAL1 & 1 & 1 & 0 \\
REF1 & 1 & 0 & 1 \\
\noalign{\smallskip}\hline
\end{tabular}
\end{table}

Although SARAS is designed to measure the spectrum of the radio background over the octave band 87.5 to 175~MHz, the lower cutoff has been set  at 110~MHz to avoid radio frequency interference from FM radio transmission below 110 MHz that is present in the Gauribidanur site where SARAS was deployed for the observations described here.

\section{Measurement equations} 

Noise voltages from the antenna and reference are denoted by $v_{a}$ and $v_{ref}$; the excess noise voltage from the calibration noise source when it is `on' is denoted by $v_{cal}$. The voltage gains along the two receiver chains from the LNA inputs up to the sampler input of the correlator are denoted by $G_{1}$ and $G_{2}$. The gain-loss in signal propagation through the cross-over switch and via the $\Sigma$ and $\Delta$ port of the power splitter to either of port $P_{1}$ or $P_{2}$ is denoted by $g$. $v_{n1}$ and $v_{n2}$ denote the noise voltages of the two LNAs (referred to the LNA inputs);  fractions $f_{1}$ and $f_{2}$ of these noise voltages are assumed to travel opposite the nominal signal flow direction.

As discussed in detail in \citet{Patra13},  signals from the antenna, calibration and reference noise sources and receiver noise suffer multiple internal reflections at impedance mismatches in the transmission path.  Impedance mismatches are usually characterized in terms of complex voltage reflection coefficients.  In the SARAS configuration, $\Gamma_{a}$ represents the mismatch at the antenna end as seen by signals propagating upwards along the 5-m cable and $\Gamma_{1}$ and $\Gamma_{2}$ represent the equivalent mismatch at the inputs of the two LNAs as seen by signals propagating down along the 5-m cable.  In this work we aim to measure the spectrum of the radio background with 1\% fractional accuracy using SARAS and, therefore, we consider only first order reflections. The $l = 5$~m  interconnecting cable between the antenna and receiver results in a roundtrip phase delay of $\Phi = (4 \pi l \nu / f_{\nu} c)$ at a frequency $\nu$, where $c$ is the velocity of light and $f_{\nu}$ is the velocity factor for propagation of EM signals in the observing frequency band in the cable. 

As discussed above, the net cross-power response in the two switch positions have opposite signs.  Differencing the measurements recorded in the two switch positions cancels any spurious additive contamination that enters the two arms of the correlation spectrometer downstream of the switch.  This difference spectrum may be written as:
\begin{equation}
P_{obs}   =  2 G_{1}G_{2}^* g^2 [v_{a}^2C_{1}  - v_{ref}^2  C_{2} + C_{n1} v_{n1}^2 +C_{n2} v_{n2}^2] 
\label{pobs}
\end{equation} 
when the calibration noise is in `off' state.  Here $2  G_{1}G_{2}^* g^{2}$ is the ideal complex gain of the correlation receiver,
\begin{equation}
 C_{1} =   1 +  g^{4} |(\Gamma_1+ \Gamma_2)\Gamma_a|^2+  2  g^{2}  {\rm Re}\left \{(\Gamma_1+ \Gamma_2)\Gamma_a e^{i\Phi} \right \},
\end{equation}
\begin{equation}
 C_{2}  =   1 - g^{2} |(\Gamma_1 - \Gamma_2)\Gamma_a|^2 + i 2 g^{2} {\rm Im} \left \{ (\Gamma_1 - \Gamma_2)\Gamma_a e^{i\Phi} \right \},
\label{c2}
\end{equation}
\begin{equation}
 C_{n1} =     \left[  g^{4}| f_1 \Gamma_a |^2  +  (f_1 \Gamma_a e^{i\Phi})^{*} \right ],
\label{cn1}
\end{equation}
and
\begin{equation}
  C_{n2} =     \left[  g^{4} | f_2 \Gamma_a |^2  +  (f_2 \Gamma_a e^{i\Phi})^{*}  \right].
\label{cn2}
\end{equation}
are the complex path gains due to multipath propagation---internal to the system---of the corresponding noise voltages. When the calibration noise source is `on', $v_{cal}^2$ adds to the antenna power $v_{a}^2$.

Differencing the spectrum recorded with the calibration noise source `on' with that recorded with the noise `off' gives the bandpass response
\begin{equation}
P_{bpc}   =  2G_{1}G_{2}^* g^{2}v_{cal}^{2} C_{1}.       
\label{pbpc}     
\end{equation}  
This is used to calibrate the spectra observed and convert to a noise temperature scale:
\begin{eqnarray}
T_{a}^{\prime} =  { P_{obs} \over P_{bpc} } \times T_{cal} 
 = \left[{v_{a}^2 C_{1}  { - v_{ref}^2  C_{2} + C_{n1} v_{n1}^2 +C_{n2} v_{n2}^2 } \over { v_{cal}^{2} C_{1}  } } \right] T_{cal}, 
\end{eqnarray} 
where $T_{cal}$ is the absolute noise temperature of the calibration noise source referred to the antenna input of the spectrometer.   In this expression $T_{a}^{\prime}$ is the antenna temperature plus additive contaminations from reference and receiver noise temperatures; we reserve the symbol $T_{a}$ to represent the antenna temperature.

The reference noise temperature is denoted by symbol $T_{ref0}$ when the reference noise source is `off', {\it i.e.}, when the reference is a termination at ambient temperature.  $T_{ref1}$ is the reference noise temperature when that noise source is `on'.  Additionally, $T_{n1}$ and $T_{n2}$ are the noise temperatures corresponding to the LNA noise voltages.  In terms of these noise temperatures,
\begin{equation}
T_{a0}^{\prime}  = T_{a} -   {C_{2} \over C_{1}} T_{ref0}
 +{C_{n1} \over C_{1}}  T_{n1} 
 +{C_{n2} \over C_{1}} T_{n2},
 \label{raw_data}
\end{equation}
and
\begin{equation}
T_{a1}^{\prime}  = T_{a} -   {C_{2} \over C_{1}} T_{ref1}
 +{C_{n1} \over C_{1}}  T_{n1} 
 +{C_{n2} \over C_{1}} T_{n2},
\end{equation} 
where $T_{a0}^{\prime}$ and $T_{a1}^{\prime}$ represent the antenna temperature plus additive contaminations at times when the reference noise source is `off' and `on' respectively.  Their difference yields a calibration product:
\begin{equation}
T_{a0}^{\prime} - T_{a1}^{\prime}  =  {C_{2} \over C_{1}} (T_{ref1}-T_{ref0})
\label{ref_diff}
\end{equation} 
that is independent of the antenna temperature as well as receiver noise temperatures.  This calibration product would manifest additive errors in both the real and imaginary components arising from multi-path propagation of the reference noise temperature and, therefore, to within a multiplying scale factor it serves as a `template' of the contribution from reference noise to the system response.


\section{Absolute calibration of the spectrometer}
\label{abs_cal}

The absolute calibration of SARAS is achieved by replacing the antenna with a matched termination, which is set to different physical temperatures.  The termination is matched to better than 50~dB and hence we may assume that noise equivalent to the physical temperature is coupled into the input cable. The physical temperatures of the matched termination and the reference termination are monitored using platinum resistance temperature probes that have 0.15~K absolute accuracy.  The matched termination is immersed separately in ice water as a `cold' load, in hot water as a `hot' load and in water at ambient temperature, and the measurements are analyzed to derive the spectral distribution of $T_{cal}$.

\begin{figure*}[ht]
\begin{minipage}[b]{0.5\linewidth}
\centering
\includegraphics[angle=270, width=0.7\linewidth]{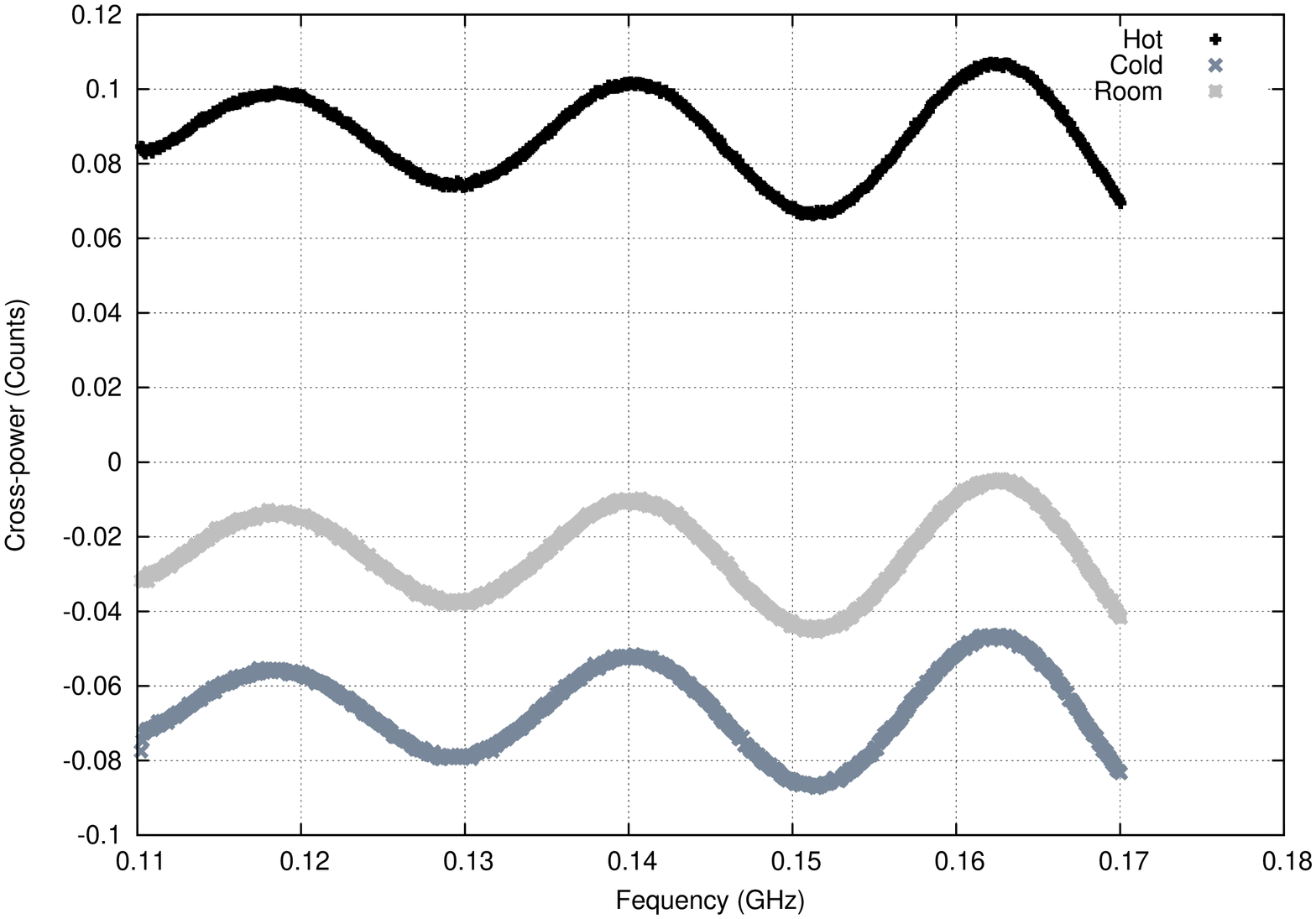}
\end{minipage}
\begin{minipage}[b]{0.5\linewidth}
\centering
\includegraphics[angle=270, width=0.7\linewidth]{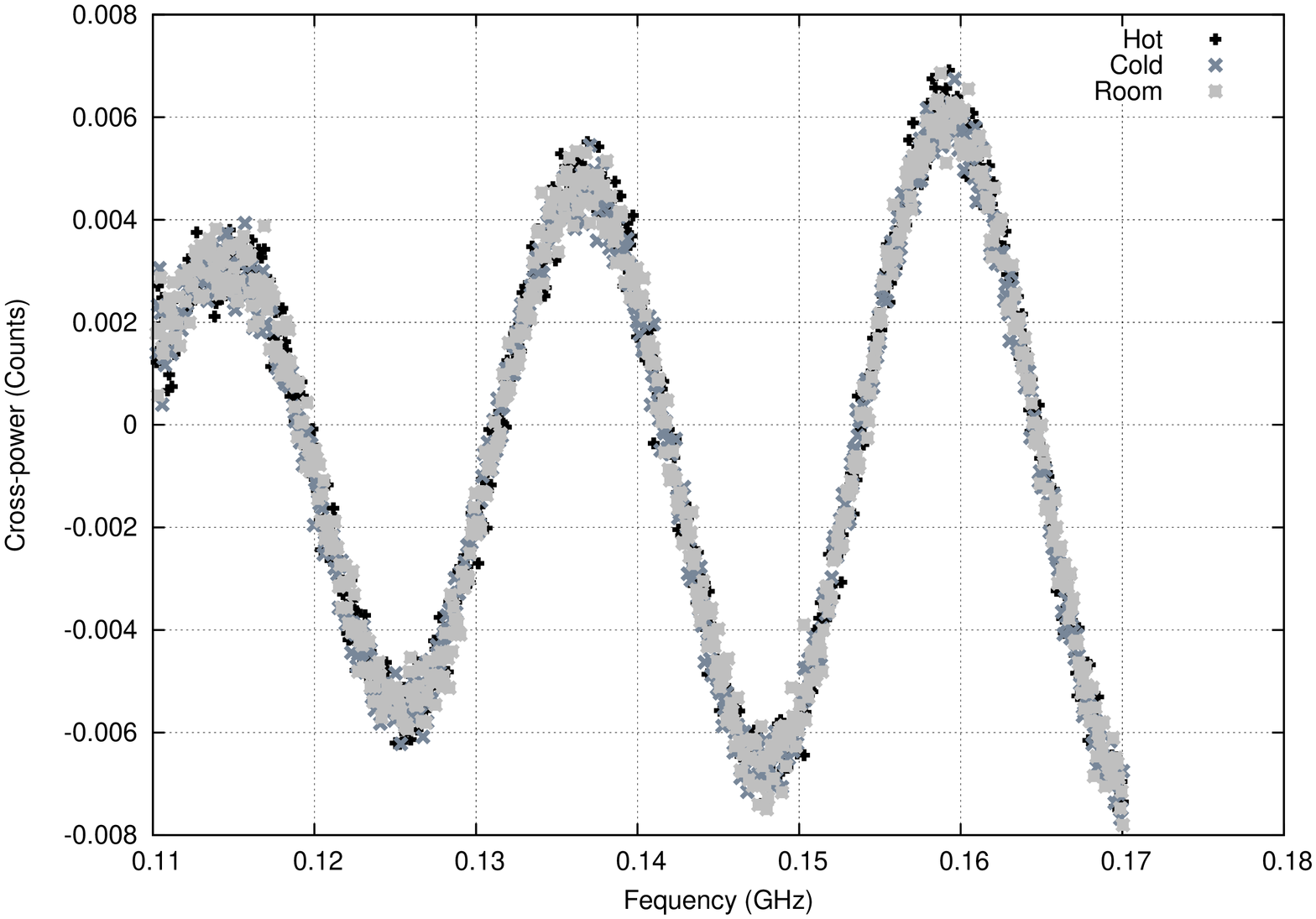}
\end{minipage}
\begin{minipage}[]{0.5\linewidth}
\centering
\includegraphics[angle=270, width=0.7\linewidth]{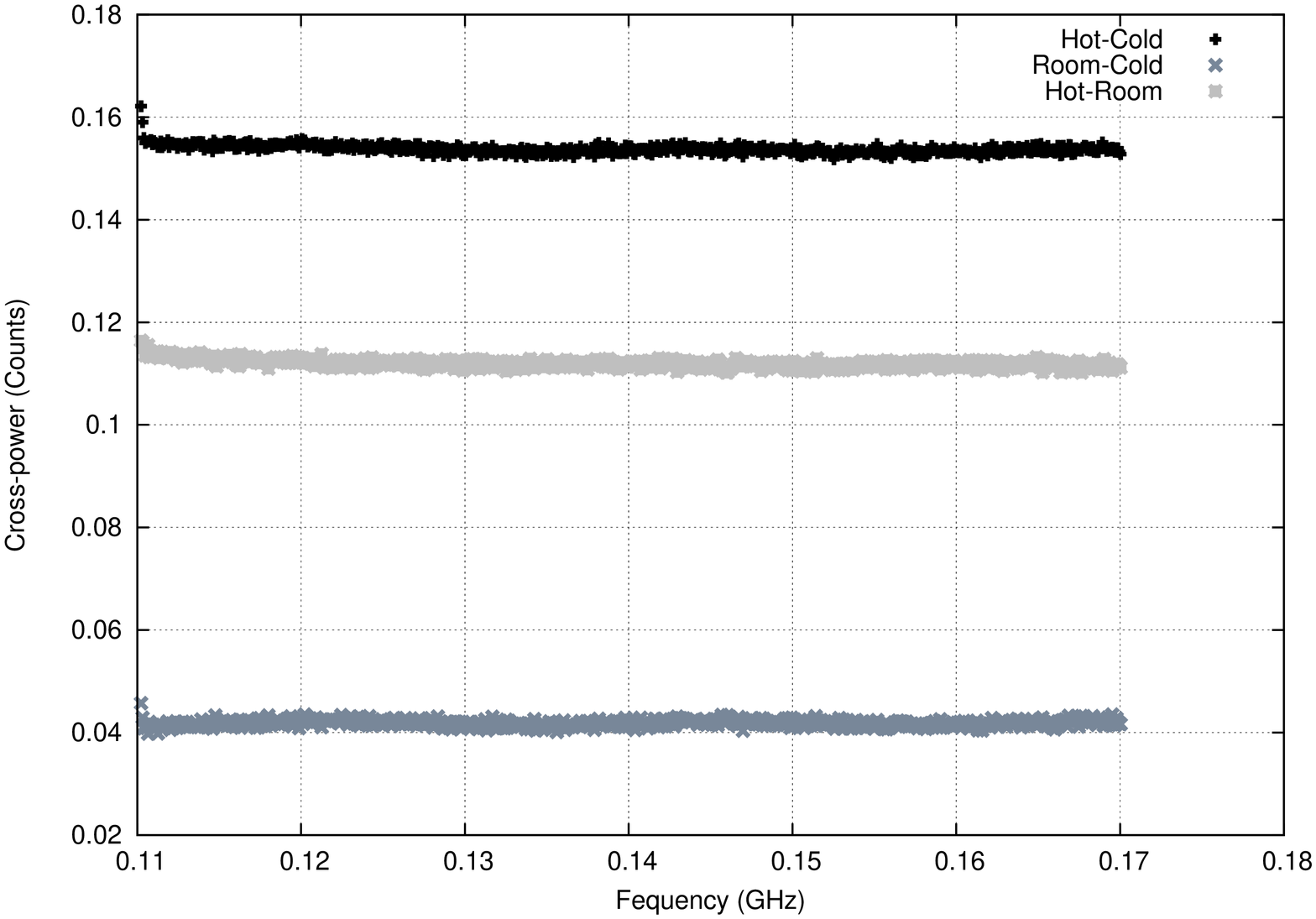}
\end{minipage}
\begin{minipage}[]{0.5\linewidth}
\centering
\includegraphics[angle=270, width=0.7\linewidth]{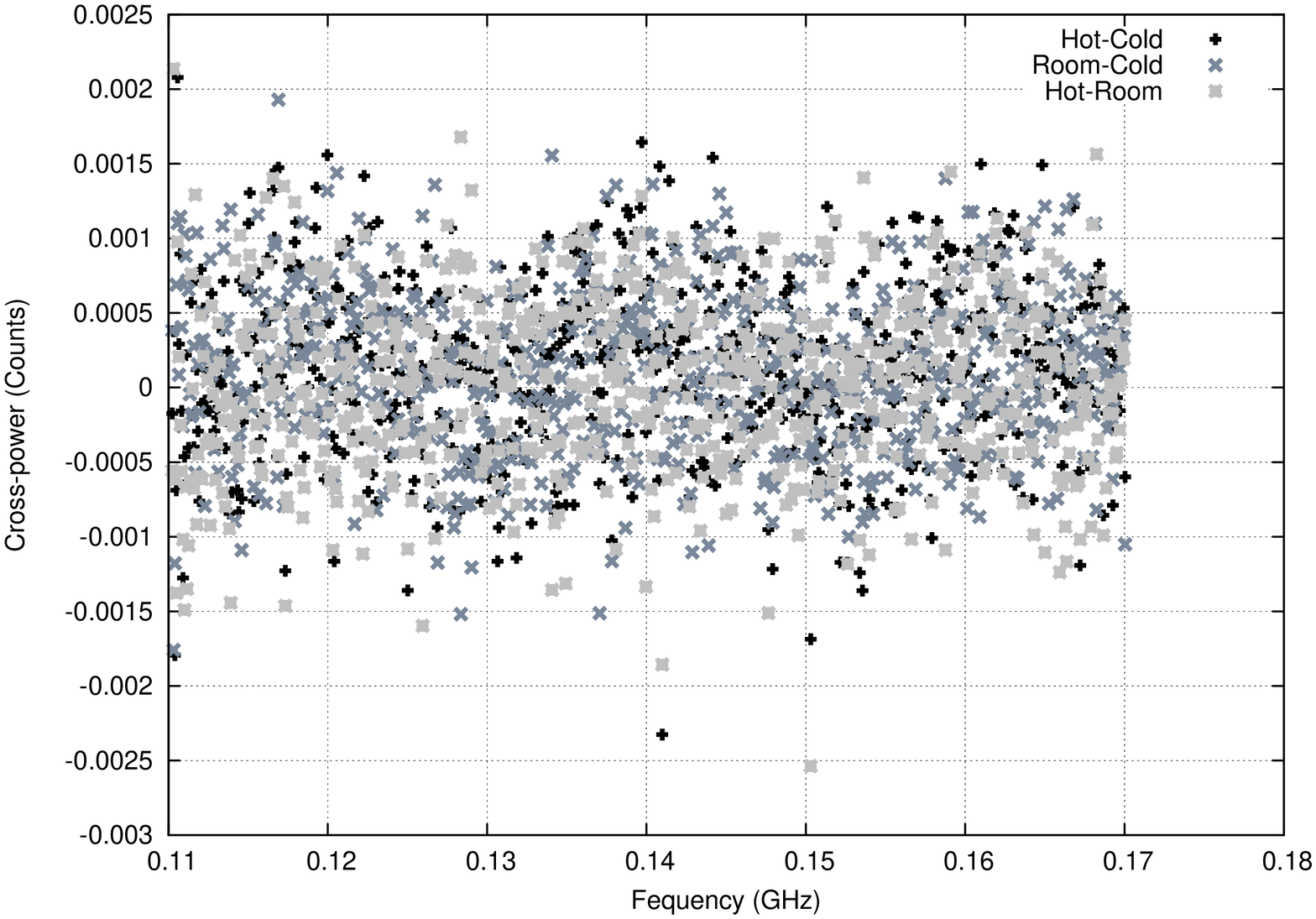}
\end{minipage}
\caption{Upper panel: Bandpass calibrated cross power spectrum measured with a matched termination in place of the antenna, which was at three different physical temperatures. Lower panel: Differences between the bandpass calibrated spectra. In both upper and lower panels, on the left is shown the real parts of the difference spectra and in the right is shown the imaginary components.}
\label{step_SARAS}
\end{figure*}

These physical temperatures are also the noise temperatures $T_{hot}$, $T_{amb}$ and $T_{cold}$, since the termination is a matched load. $P_{obs}$ spectra recorded with the termination at three different temperatures are referred to as $P_{hot}$, $P_{amb}$ and $P_{cold}$ respectively.  The recorded complex cross spectra were calibrated assuming a $T_{cal}$ of unity and the computed $(P_{hot}/P_{bpc})$, $(P_{amb}/P_{bpc})$ and $(P_{cold}/P_{bpc})$ spectra for the three load temperatures are shown in Fig.~\ref{step_SARAS}.  The ripples in the recorded spectra are because of the mismatch seen by 
noise from the reference termination and low noise amplifiers as they propagate towards the matched termination that is set to different physical temperatures and the consequent multipath propagation with roundtrip phase delay of $\Phi = (4 \pi l \nu / f_{\nu} c)$.  The ripple amplitude is consistent with the measured reflection coefficient between the 5-m cable and the direction coupler that injects that calibration noise into the system path.  


The mean temperature of the reference $T_{ref0}$ was  303~K when the load was `hot' and 302.5~K when `cold', and varied by less than 0.5~K during the calibrations. The LNA noise temperatures are constant during calibration and since the reference noise temperature is also unchanging to within a fraction of a per cent, the amplitudes of the observed ripples in the calibration data acquired with different load temperatures are same.  

The lower panels in Fig.~\ref{step_SARAS} show the differential calibration spectra: $[(P_{hot}/P_{bpc}) - (P_{cold}/P_{bpc})]$, $[(P_{amb}/P_{bpc}) - (P_{cold}/P_{bpc})]$ and $[(P_{hot}/P_{bpc}) - (P_{amb}/P_{bpc})]$.  These spectra have flat real components and the imaginary components are consistent with zero response. The real components are proportional to the differences $T_{hot}-T_{cold}$,  $T_{amb}-T_{cold}$ and $T_{hot}-T_{amb}$, with a correction for the change in $T_{ref0}$ during the pairs of measurements in each case. 
Denoting the reference temperatures as $T_{ref0\ hot}$ and $T_{ref0\ cold}$ when the load replacing the antenna is in `hot' and `cold' states, we may write:
\begin{equation}
T_{cal} = { \left \{(T_{hot}-T_{ref0\ hot}) - (T_{cold}-T_{ref0\ cold})\right \} \over \left( {P_{hot}\over P_{bpc}} - {P_{cold} \over P_{bpc}} \right)}.
\label{Tcal}
\end{equation}


For a flat calibration noise source, the measured $T_{cal}$ spectrum derived here would be a constant; our calibration measurement shows that this is indeed true; the calibration data yields a constant $T_{cal}$ value of 503.3~K with fractional error of 0.6\%.


\section{Observations and data analysis}


The SARAS spectral-line radiometer was located at the Gouribidanur radio observatory (latitude: $13\fdg6$ N and longitude: $77\fdg44$ E) that is about 80~km north of Bangalore, India.  The center of the horizontal fat-dipole antenna was placed 0.8~m above the ground, which was covered by  flat absorber tiles made of Nickel-Zinc Ferrite composition (Panashield SFA-type) that has reflectivity as low as $-30$~dB across the observing band.  Measurement sets were recorded with 0.7~s integration times sequentially in each of the six states listed in Table~\ref{table:states} and this sequence of observations were repeated for the entire observing session.  The observation was done during LST 23$^h$ to 01$^h$.  From observations in the LST range 23$^{h}$ to 1$^{h}$, 20 calibrated spectra are obtained by averaging spectra within 6~min blocks.  


\subsection{The derived data and calibration products}

\begin{figure*}[ht]
\begin{minipage}[b]{0.5\linewidth}
\centering
\includegraphics[angle=270, width=0.7\linewidth]{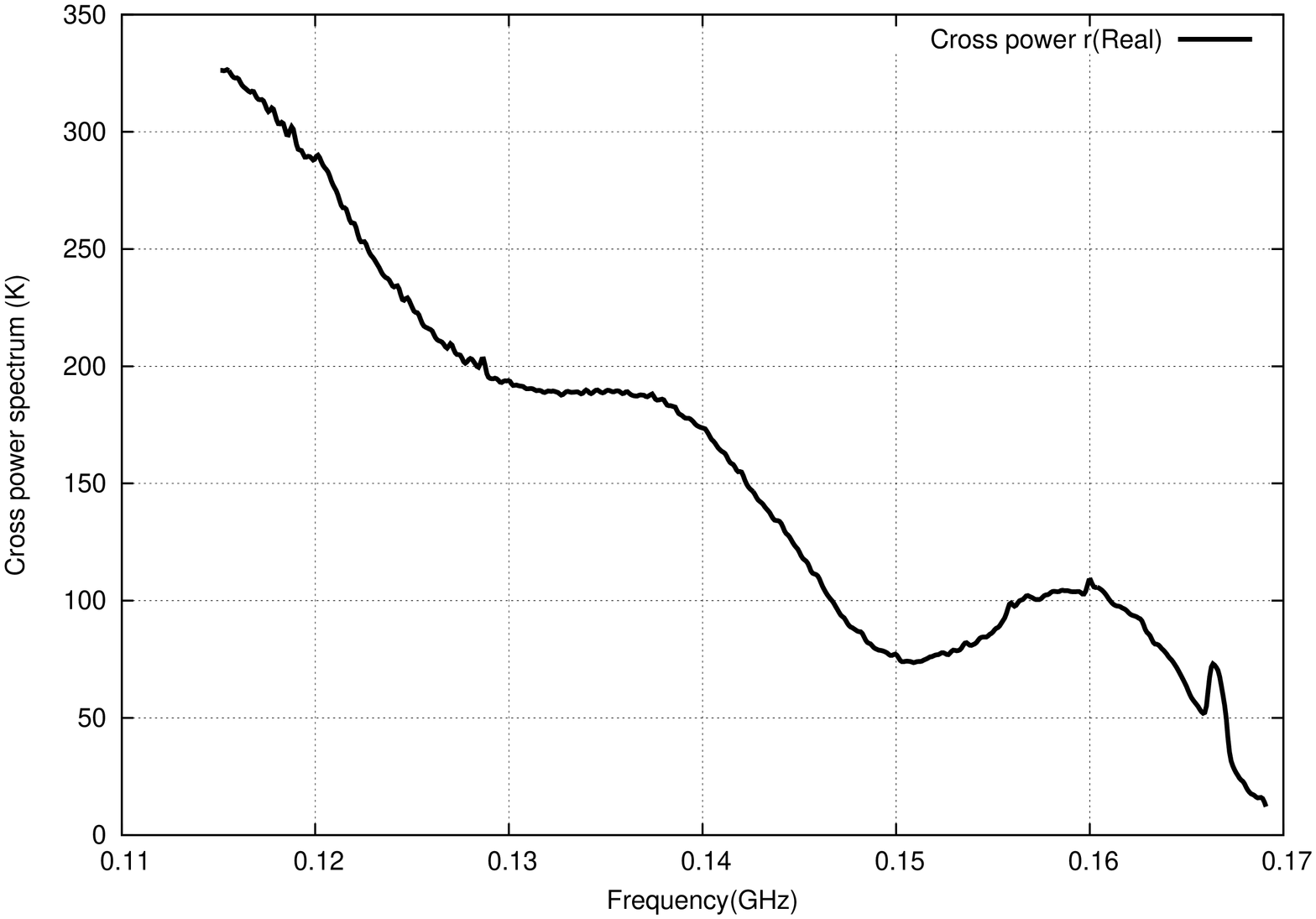}
\end{minipage}
\begin{minipage}[b]{0.5\linewidth}
\centering
\includegraphics[angle=270, width=0.7\linewidth]{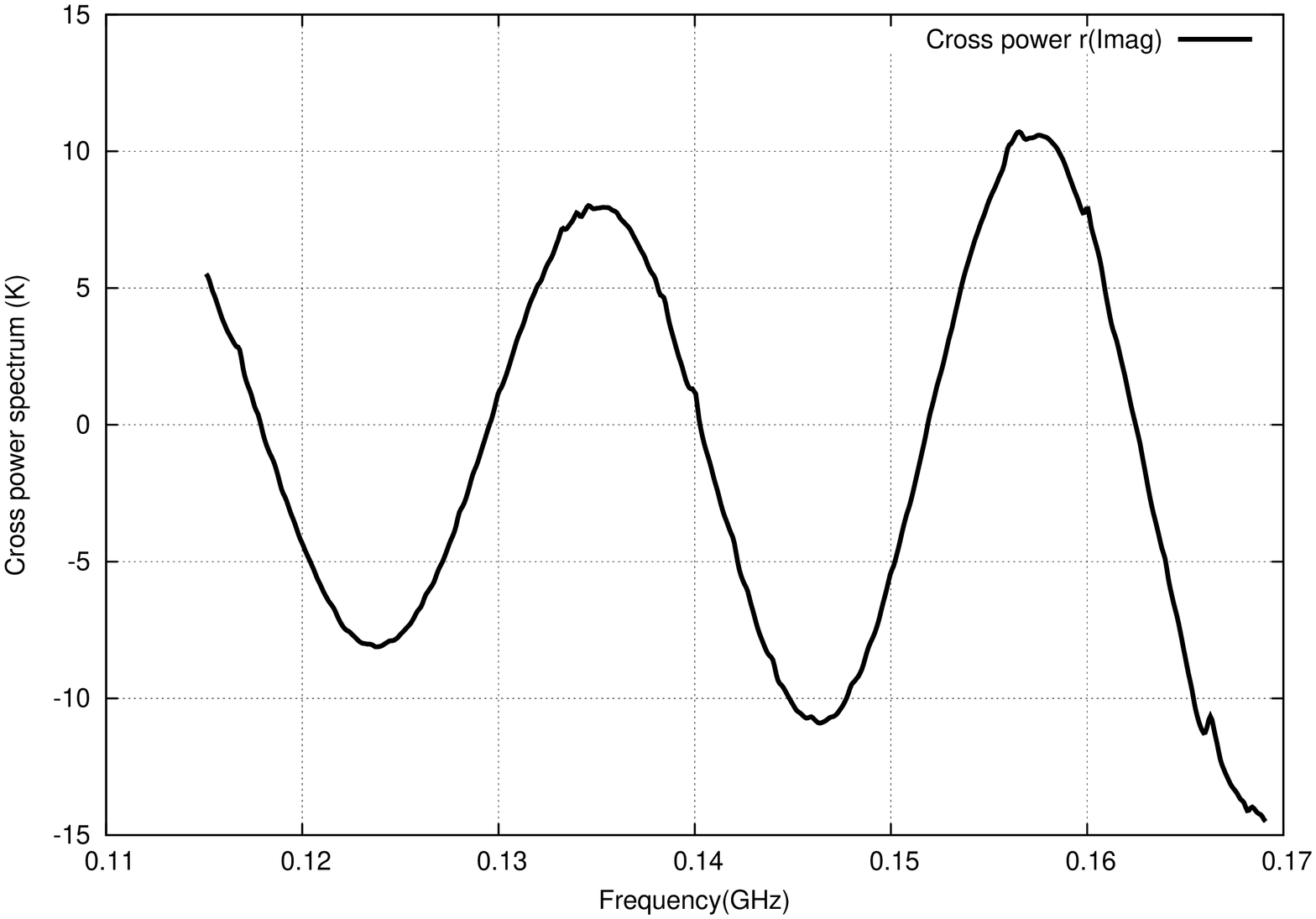}
\end{minipage}
\begin{minipage}[b]{0.5\linewidth}
\centering
\includegraphics[angle=270, width=0.7\linewidth]{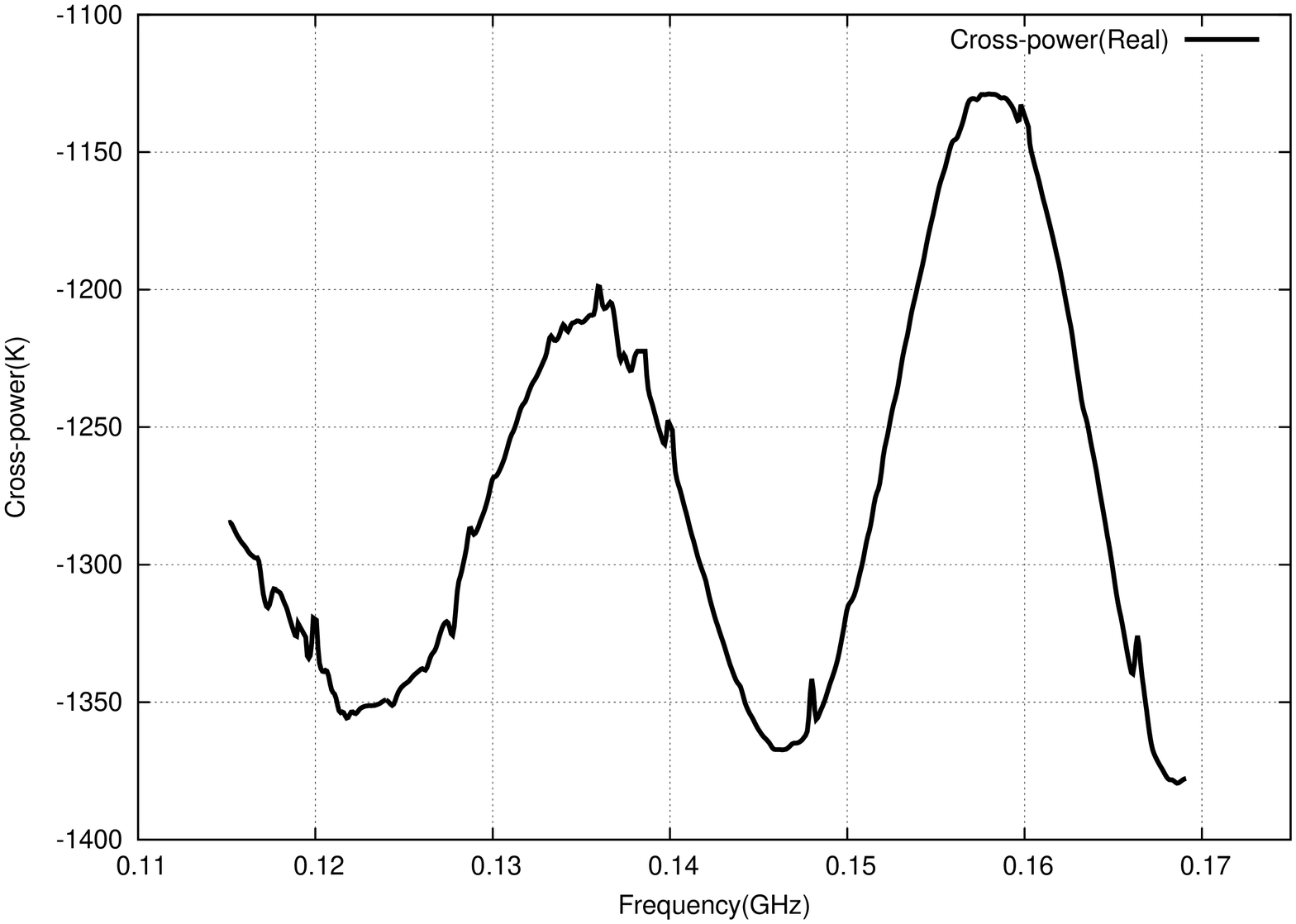}
\end{minipage}
\begin{minipage}[b]{0.5\linewidth}
\centering
\includegraphics[angle=270, width=0.7\linewidth]{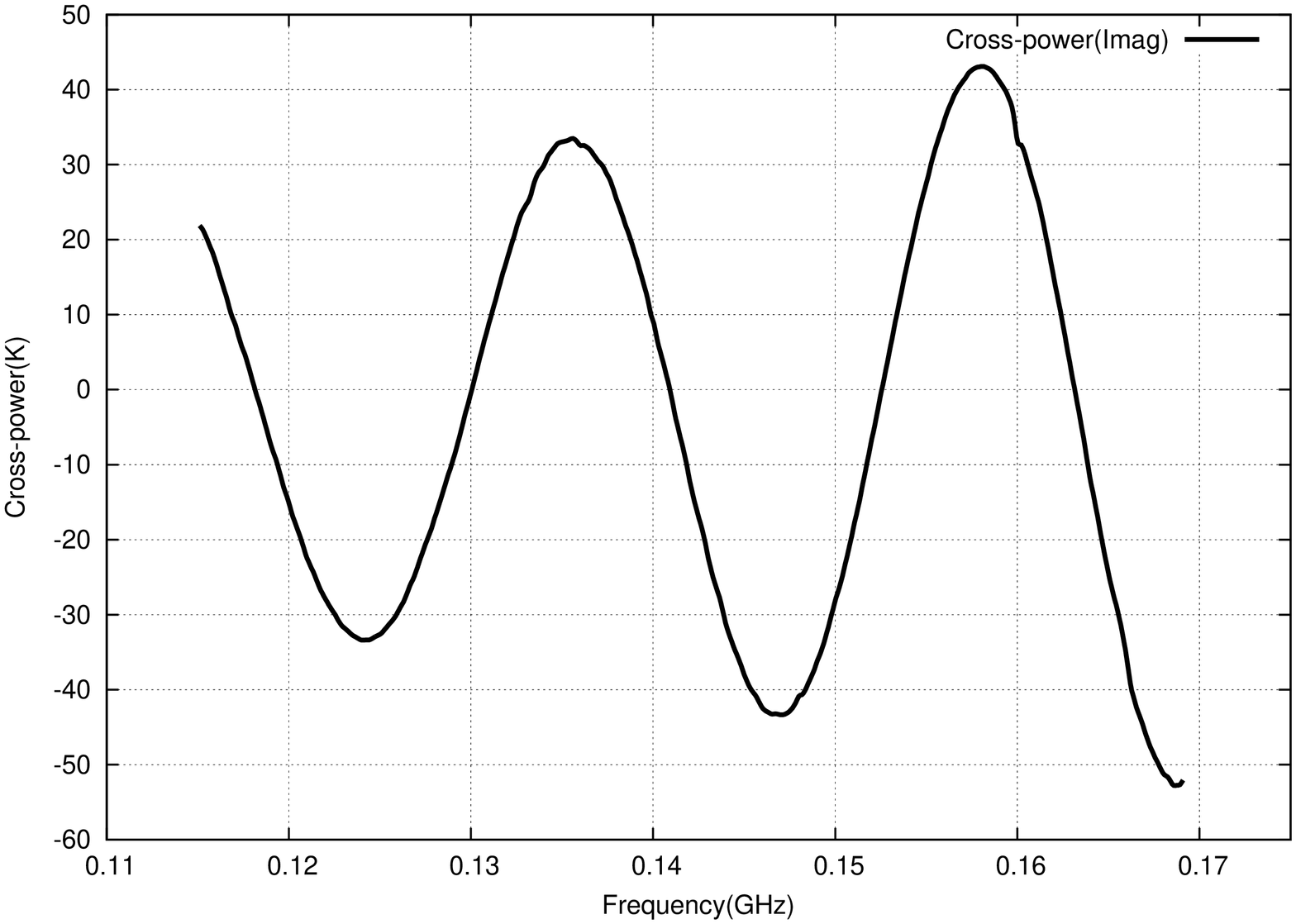}
\end{minipage}
\caption{Upper panels: Example of a calibrated cross-power spectrum measurement made with the reference noise source `off'. Lower panels: Difference between the calibrated spectra computed with reference in `on' and `off' states: this complex spectrum is a template of the response to reference noise.  The panels on the left and right show, respectively, the real and the imaginary parts. }
\label{hot_SARAS}
\end{figure*}

For each set of six complex cross-correlation spectra recorded every 4.2~s, the bandpass response of Equation~6 was derived and thence the calibrated measurement set corresponding to Equation~8 was computed.  Additionally, the calibration product corresponding to Equation~10, which is a template of the response to reference noise, was computed.  As an example, in Fig.~\ref{hot_SARAS} we show  the calibrated complex measurement set (which was obtained with reference noise `off') and the calibration product that is a measure of the response to reference noise. The real part of the former has contributions from the antenna, reference and receiver noise temperatures where as the imaginary consists only of reference and receiver temperature contributions.   The calibration product represents response to reference noise and has no contribution from either the antenna or receiver noise.  


\subsection{Modeling the SARAS data}
\label{Modelling}

The primary data product of the `SARAS' system, which is the real part of the calibrated measurement set defined by Equation~ \ref{raw_data}, consists of three components:  the antenna temperature $T_{a}$, the reference noise temperature $T_{ref0}$ and the two receiver noises $T_{n1}$ and $T_{n2}$.  Modeling involves a joint fit to this data set and to the imaginary part of the calibrated measurement set.  Used in the modeling are the derived calibration product---template of the response to reference noise---and other calibration products based on laboratory measurements on the antenna and its balun.

\subsubsection{A model for the antenna temperature}

SARAS antenna has the radiation pattern of a short dipole. Half the beam solid angle of the dipole observes the sky whereas the other half receives  ground radiation. Therefore, the antenna temperature  $T_{a} $ is the beam-weighted average brightness temperature of the sky $T_{sky}$ and the ground temperature $T_{gnd}$. $T_{gnd}$ is measured by a platinum thermometer. We  model the radio background as a single power law of the form $T_{RB} = T_{0}({\nu \over \nu_{0}})^{\alpha}$,  which has two free parameters $T_{0}$ and $\alpha$ to be estimated.  Since the voltage return loss at the antenna terminal is $\Gamma_{a}$, the net external noise power coupled to the system by the antenna at the balun output is ${ \left(T_{RB}+T_{gnd} \over 2 \right) }(1-|\Gamma_{a}|^2)g_{bl}$  where $g_{bl}$ is the balun resistive loss. We assume that the antenna has no resistive loss and it is only the balun that adds a noise temperature $(1-g_{bl})T_{amb}$.  Both the antenna return loss and balun resistive loss are measured using test and measurement instruments and are not parameters to be solved for.  Since the SARAS antenna is a frequency independent short dipole, the antenna directivity is a constant over frequency. Therefore, the antenna temperature $T_{a}$ in Equation \ref{raw_data} is modeled as:
\begin{equation}
 T_{a} = \left [{\left(T_{0}({\nu \over \nu_{0}})^{\alpha} + T_{gnd}\over 2\right)} \right] (1-|\Gamma_{a}|^{2})g_{bl} + (1-g_{bl})T_{amb}.
 \label{Tant}
 \end{equation}
 
\subsubsection{Reference and receiver noise model}

The reference and receiver noise contributions to the measured data are scaled by factors $C_{1}$, $C_{2}$, $C_{n1}$, $C_{n2}$ (see Equation~\ref{raw_data}), which are functions of complex system parameters $\Gamma_{a}$, $\Gamma_{1}$ and $\Gamma_{2}$.  

A platinum thermometer measures the reference temperature $T_{ref0}$.  Most importantly, the derived calibration product $T_{a0}^{'}-T_{a1}^{'}$ yields the system response to the excess reference noise $(T_{ref1}-T_{ref0})$.  Therefore, this serves as a template for modeling the reference noise contribution to the primary data set (Equation~\ref{raw_data}), and will represent the true response to noise from the reference termination after scaling by a factor $S$ that is the ratio $\{T_{ref0}/(T_{ref1}-T_{ref0})\}$.  It may be noted here that this scaled template accounts for all the higher order reflections of the reference noise as well.

The response to receiver noise is dominated by the second terms in Equations~\ref{cn1} and \ref{cn2}, which dominate the leading terms.  These are expected to appear in the real and the imaginary components of the measurement set as zero mean quasi-sinusoids that are in approximate phase quadrature and with unequal amplitudes.  If the fractions of the receiver noise $f_{1}$ and $f_{2}$ that travel upstream in the signal path are complex then this may result in an additional phase difference between the real and imaginary parts as well as a small difference in frequency.  We therefore model the receiver noise contribution to the real and the imaginary components of the measurement set respectively by sinusoids of frequency and frequency offset $w$, $\Delta w$ that have a relative phase $\Delta \phi$ and which are amplitude modulated by functions $T_{nr}(\nu)$, $T_{nm}(\nu)$. \\

From Equation~\ref{raw_data}, receiver noise contribution may be written in the form: 
\begin{eqnarray}
{C_{n1} \over C_{1}}T_{n1}+{C_{n2} \over C_{1}}T_{n2} & = &  |\Gamma_a| [ {\left( f_1 V_{n1}^2 + f_2 V_{n2}^2  \over C_{1} \right) } {\rm \cos} (\Phi + \Phi_a)  - i {\left( f_1 V_{n1}^2 - f_2 V_{n2}^2 \over C_{1}\right)} {\rm \sin} (\Phi + \Phi_a) ]. \nonumber \\
& = &  T_{nr}(\nu)\ \cos \left[(w+\Delta w)\nu + \phi_{a}\right ] - i  \ T_{nm}(\nu)\ \sin \left[(w-\Delta w)\nu+ \phi_{a}+ \Delta \phi \right]. 
\label{eq5.3}
\end{eqnarray}
From Equation \ref{c2}, the real part of the expression for the function $C_{2}$ may be approximated as unity. The real part of Equation \ref{ref_diff} may then be used to derive an approximation for $C_1$:
\begin{equation}
C_1 = \Delta T_{ref} / [{\rm Re} (T_{a0}^{'} - T_{a1}^{' })],
\end{equation}
where $\Delta T_{ref} = (T_{ref1}-T_{ref0})$. The modulating functions $T_{nr}(\nu)$ and $T_{nm}(\nu)$ may then be expressed in terms of the real component of the calibration product template and the antenna reflection coefficient as:
\begin{equation}
T_{nr}(\nu)= t_{nr} |\Gamma_{a}| {\rm Re} (T_{a0}^{'} - T_{a1}^{' })
\label{Tnr}
\end{equation}
and
\begin{equation}
T_{nm}(\nu)= t_{nm} |\Gamma_{a}| {\rm Re} (T_{a0}^{'} - T_{a1}^{' }). 
\label{Tnm}
\end{equation}
Factors $t_{nr}$ and $t_{nm}$  are model parameters along with frequency and frequency offset $w$, $\Delta w$ and relative phase $\Delta \phi$ that together describe the response of the system to receiver noise. Unlike the more precise modeling of the reference noise, only the first order reflections are taken into account in modeling the receiver noise and this has been deemed adequate considering the accuracy we aim for.

In summary, the measurement data of SARAS is modeled by a set of eight parameters: (i) model parameters for $T_{sky}$: $T_{0}$ and $\alpha$, (ii) a scaling factor for the reference noise template: $S$ and (iii) receiver noise parameters: $t_{nr}$, $t_{nm}$, $w$, $\Delta w$, and $\Delta \phi$.
 Initial guess for the system parameters are derived from the measurement data without relying on laboratory measurements.  The complete model that we adopt for describing the real and imaginary parts of the calibrated measurement set is:
 \begin{eqnarray}
{\rm Re}[T^{'}_{a0}|_{\rm model}] & = &  \left(T_{0}({\nu \over \nu_{0}})^{\alpha} + T_{gnd}\over 2\right)  (1-|\Gamma_{a}|^{2})g_{bl} + (1-g_{bl})T_{amb} \nonumber \\ 
&& + S\ {\rm Re}(T_{a0}^{'} - T_{a1}^{'})+ t_{nr} |\Gamma_{a}| {\rm Re} (T_{a0}^{'} - T_{a1}^{' }) \cos \left[(w+\Delta w)\nu + \phi_{a}\right ]
\label{real_model}
 \end{eqnarray} 
 and
\begin{equation}
 {\rm Im}[T^{'}_{a0}|_{\rm model}] = S\ {\rm Im}(T_{a0}^{'} - T_{a1}^{'})+ t_{nm} |\Gamma_{a}|  {\rm Re} (T_{a0}^{'} - T_{a1}^{' }) \sin \left[(w-\Delta w)\nu+ \phi_{a}+ \Delta \phi \right].
 \label{imag_model}
 \end{equation}
  This model is jointly fitted to the real and imaginary components of the primary data product, which is the calibrated measurement set (Equation~\ref{raw_data}).

 \subsection{Hierarchical model fitting}

Given a calibrated spectrum at any observing time, we compute the best-fit values for these eight parameters and the likelihood distributions of  $T_{0}$ and $\alpha$. In other words, given a measured data set D, we first determine the posterior probability distribution $P(\Theta | D)$ where $\Theta$ is an eight dimensional vector of model parameters. Marginalizing the posterior probability distribution function over the system parameters will then result in a two dimensional probability distribution of the parameters $T_{0}$ and $\alpha$ that describe the sky spectrum.  We derive the best fit values of the parameters by minimizing the merit function ${\chi}^{2}$:
\begin{equation}
{\chi}^{2} = \displaystyle\sum\limits_{i=0}^{n}\left[{T_{model}(\nu_{i}) - T_{obs}(\nu_{i})\over 2{\sigma}^2(\nu_{i})}\right]^2.
\end{equation} 
The merit function has non-linear dependence on the model parameters; therefore, chi-square is not unimodal with a single minimum. Hence, it is extremely important to start any ${\chi}^{2}$ minimization with good initial guess for each parameter and use  appropriate priors. Since all sources of noise (antenna, reference and receiver) are uncorrelated, a hierarchical approach may be adopted to model individual noise contributions in which parameters corresponding to dominant contributors are first estimated.  

\begin{enumerate}
\item 
\textbf{Step 1:} First, the real part of the calibrated measurement set is fitted to 
\begin{equation}
{\rm Re}[T^{'}_{a0}|_{\rm model}] =  \left(T_{0}({\nu \over \nu_{0}})^{\alpha} + T_{gnd}\over 2\right)  (1-|\Gamma_{a}|^{2})g_{bl} + (1-g_{bl})T_{gnd} - T_{ref0}.
\label{step1_mod}
\end{equation} 
Thermometer measurements of ground temperature are assumed for balun ohmic loss and ground emission; the thermometer measurement of the reference termination is assumed for $T_{ref0}$.  

Two parameter model ($T_{0}$ and $\alpha$) for the average sky brightness in the antenna beam is solved for.  Initial guess for these parameters is based on previous measurements of the radio background.  The SARAS band lies at the upper end of the frequency range covered by previous measurements by \cite{Costain60} who estimated the spectral index to be  $-2.37\pm0.04$; higher frequency measurements (summarized in Section~1) suggest a steepening there and hence 
we initialise the value of  $\alpha$  to $-2.5$.   The measurement of \cite{Costain60} suggests the mean sky brightness temperature between RA 16 to 2$^{h}$ is about 250~K at 178~MHz. For the adopted spectral index of $-2.5$, the temperature  at 150~MHz is estimated to be about 400~K and this guess is used for $T_{0}$. 
For the 20 time-averaged spectra recorded in the observed LST range, the estimated spectral index and temperature $T_{0}$ are shown in Figure~\ref{Step3_param_LST}.

\begin{figure*}[ht]
\centering
\includegraphics[ angle=270,width=0.8\linewidth,totalheight=6cm]{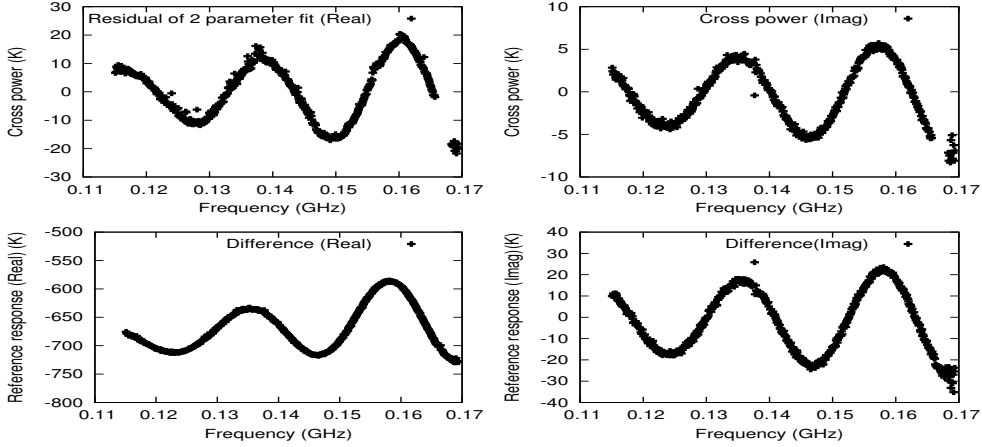}
\caption{Upper panel: Residual after the fit to the real part of the primary spectrum based on the model in Step~1. Lower panel: Template of the system response to  reference noise.}
\label{step1_resid}       
\end{figure*}

\item 
\textbf{Step 2:} 
In Fig.~\ref{step1_resid} is shown an example of the residual after subtracting the best fit model from the step above; this residual is dominated in its structure by the response to noise from the reference termination, the template for this response is reproduced in the lower panel of the figure for comparison.  In this second step, in addition to the parameters describing the sky brightness, the modeling of the measurement set is expanded to include the scaled template for the response to reference noise and now the real and imaginary parts of the measurement set are jointly modeled. 

 
The initial guess for the scale factor $S= T_{ref0}/\Delta T_{ref}$ is derived as the median of the ratio of the residual to the template over the observing time; this is computed separately for the real and imaginary components of the measurement set. We found that these two estimates differ and vary over time by less than about $1\%$.

\item 
\textbf{Step 3:}  In the final step, receiver noise models of the form $T_{nr}(\nu) \cos \left [(w-\Delta w)\nu+\phi_{a} \right ]$ for the real part and 
$T_{nm}(\nu) \sin \left [(w+\Delta w)\nu+\phi_{a}+\Delta \phi \right]$ for imaginary are included while modeling the measurement set.  The functions $T_{nr}(\nu)$ and $T_{nm}(\nu)$ are given by Equations~\ref{Tnr} and \ref{Tnm}.

Initial guess for the parameters $t_{nr}$ and $t_{nm}$ are obtained by dividing the residuals of Step~2 by the antenna return loss $|\Gamma_{a}|$ and real part of the difference spectrum  ${\rm Re} (T_{a0}^{'} - T_{a1}^{' })$, and computing the amplitude of the sinusoid. This yields initial guess values of about 0.06 and 0.002 for $t_{nr}$ and $t_{nm}$. 
The ripple frequency $\omega$ is estimated from the cable length and cable velocity factor whereas the frequency and phase offsets $\Delta \omega$ and $\Delta \phi$ are estimated by fitting sinusiodal functions to the residuals of Step~2. 
These values, along with the best fit values of $\alpha$, $T_{0}$ and $S$ obtained from the modeling of Step~2, are used as the initial guess and the calibrated measurement set is modeled using the eight parameter model given by Equations~\ref{real_model} and~\ref{imag_model}.
 An example of this joint fit and residuals to the fit are shown in Figure~\ref{step3}. The rms residuals to the fits to the real and imaginary  parts  are 1.15~K and  0.72~K respectively. Figure~\ref{Step3_param_LST} shows the variation of spectral index and $T_{0}$ from this joint fit over the observing time. 
 



\begin{figure*}[!htb]
\centering
\includegraphics[ angle=270,width=0.7\linewidth,totalheight=6cm]{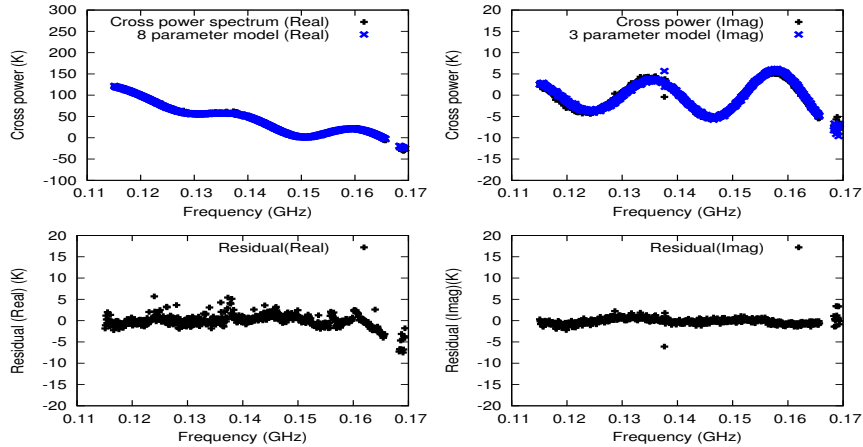}
\caption{An example data and model for the 8-parameter joint fit to the real and imaginary parts of the spectra (upper panel) and the fit residuals (lower panel). The data and model overlap and are indistinguishable in the plot. }
\label{step3}       
\end{figure*}

\begin{figure}
\begin{minipage}[b]{\linewidth}
\centering
\includegraphics[angle=270, width=0.7\linewidth]{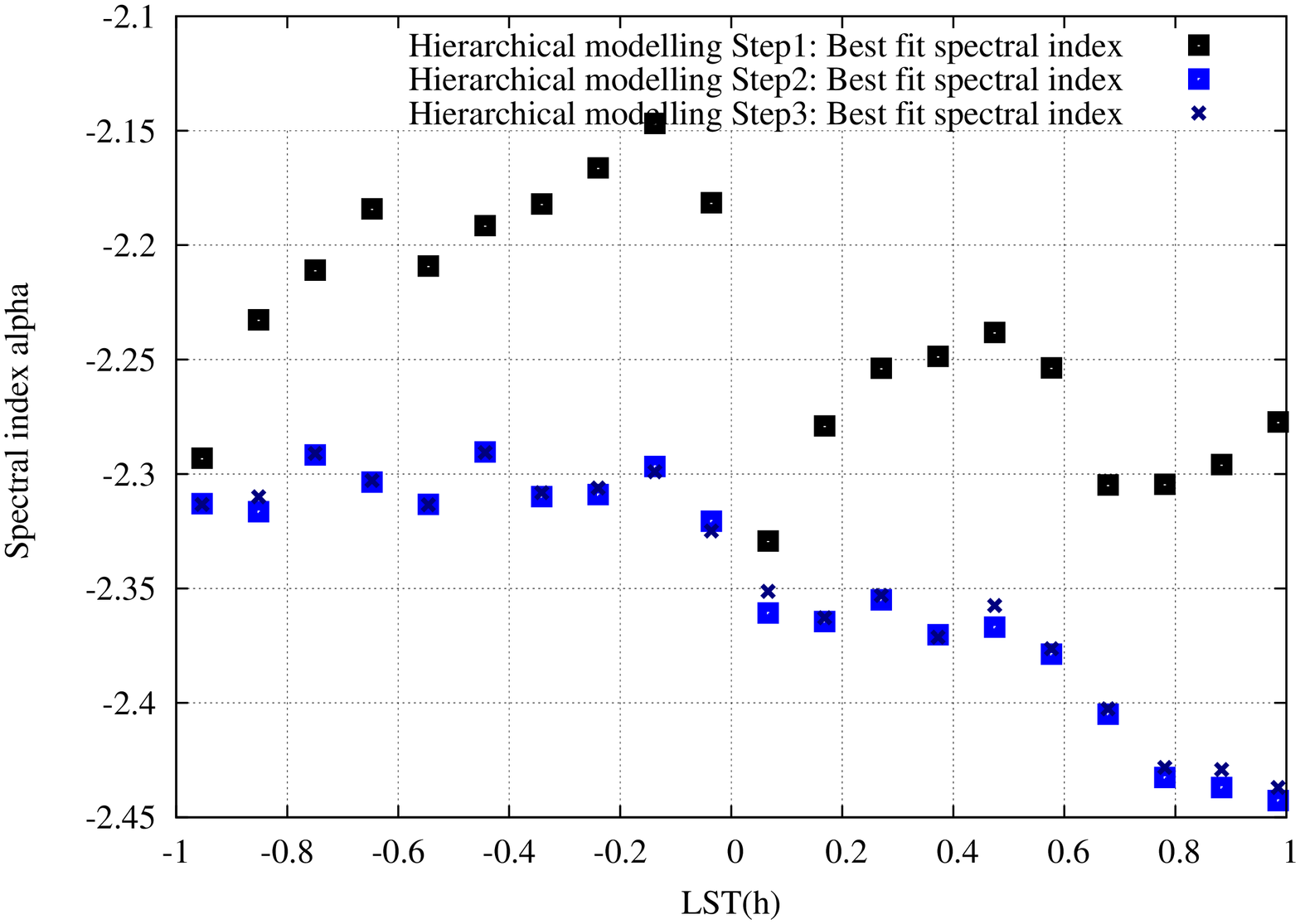}
\end{minipage}
\begin{minipage}[b]{\linewidth}
\centering
\includegraphics[angle=270, width=0.7\linewidth ]{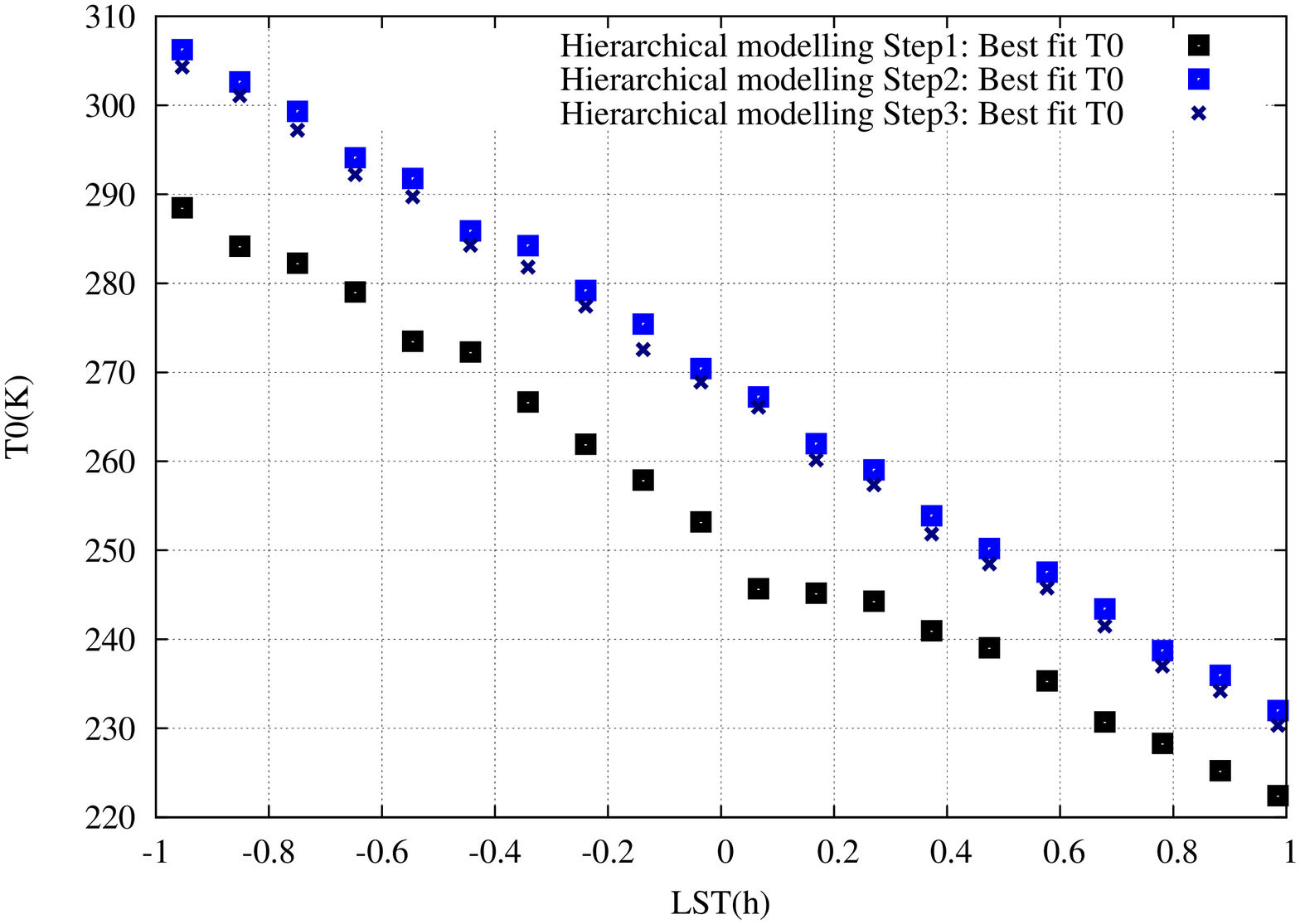}
\end{minipage}
\caption{Variation of the spectral index $\alpha$ and $T_{0}$ as a function of LST for fits to a sample spectrum. The parameters obtained from Step~3 are compared with those from Step~2 and Step~1 of the hierarchical modeling.}
\label{Step3_param_LST}       
\end{figure}
\end{enumerate}

\subsection[Measurement noise and Systematic errors]{Measurement  noise and systematic errors}
 \label{meas_sys_noise}

\subsubsection{Measurement noise}
\label{meas_noise}

The SARAS spectrometer measures the sky brightness between 87.5 to 175~MHz with the spectral resolution and channel bandwidth $\Delta \nu$ of 85.45~kHz. The mean channel power is from antenna noise, reference noise and receiver noise.   With calibration and reference noise sources both in off state, the variance in measured power in the real and imaginary parts of the correlation spectrum  is:
\begin{equation}
\Delta T_{\rm 0off}  =  \Delta T_{\rm 1off} = \sqrt{ \left [ (5/4)(T_{a}+T_{\rm ref})^{2} +(T_{a}+T_{ref}) (T_{\rm n1}+ T_{\rm n2}) + T_{\rm n1}T_{\rm n2} \right ] \over 2 \Delta \nu \Delta t}.
\end{equation}
The basic integration time $\Delta t$ is 0.7 sec.  Assuming reference noise $T_{\rm ref} = 300$~K at ambient temperature, receiver noise temperature $T_{\rm n1} = T_{\rm n2} =115$~K and sky brightness $T_{a} = 350$~K at 150~MHz (as the average of a $T_a$ of 400~K and ground brightness of 300~K), the rms variation in the channel data of bandpass calibrated spectra at 150~MHz is $2.4$~K. 
Averaging the measured power in two switch positions results in a variance that is $\sqrt{2}$ times smaller.  Since each spectrum for which a model is fitted is produced by averaging 72 calibrated spectra over 6~min intervals, the expected measurement noise in channel data in each spectrum is 200~mK.




The noise in the measured spectra is also estimated from the  data by computing rms of the differences between adjacent channel data (also called Allen variance).  
This estimated variance is 620~mK, which is larger than the expected thermal noise. 
This is believed to be owing to interference that has survived the RFI excision algorithm.
The Hierarchical modeling of the SARAS data is done in two passes. In the first pass,  the measurement error per channel is assumed to be an arbitrary constant.  The rms noise is then estimated from the residuals and this is used to redo the modeling and refine the estimation of model parameters. 

\subsubsection{Systematic errors}
 
\begin{figure}[ht]
\centering
\includegraphics[angle=270, width=0.5\linewidth]{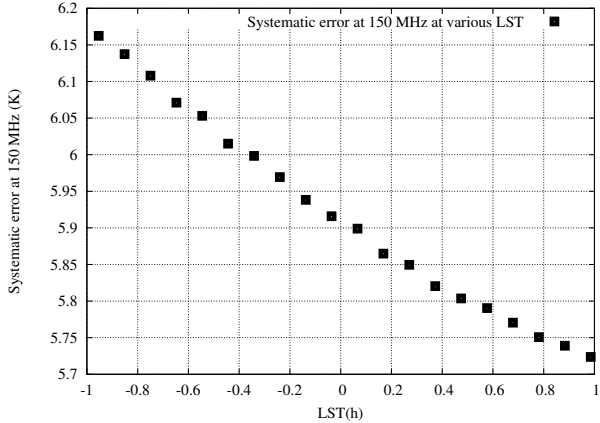}
\caption{Systematic error in the estimation of $T_{0}$, i.e., the sky brightness temperature at 150~MHz, over the LST range of the observations. The errors are estimated using the best fit values of $\alpha$ and $T_{0}$. }
\label{syser_spect}       
\end{figure}

The primary sources of systematic errors in the SARAS analysis are those associated with measurements of antenna reflection coefficient, the ground temperature and the balun resistive loss. The net systematic error may be estimated by 
linearizing the measured temperature in each case, e.g. Equation~(\ref{step1_mod}). The total systematic error is given by: 
\begin{equation}
 \Delta T_a = \left [({\delta{T_{a0}} \over \delta |\Gamma_a| }  \Delta |\Gamma_a| )^{2} +( {\delta{T_{a0}} \over \delta{T_{gnd}} }  \Delta T_{gnd})^{2} + ({\delta{T_{a0}} \over \delta g_{bl} }  \Delta g_{bl} )^{2} \right ]^{1/2}. \nonumber 
 \end{equation} 
 The measurement accuracy of the platinum thermometer that measures the ground temperature is $\pm 0.15~$C. The vector network analyser used for the antenna return loss measurement has a quoted calibration error of $0.1~$dB. In the modeling of each spectrum, the total systematic error at each frequency is calculated using the best-fit values of $\alpha$ and $T_{0}$. 
 The variation in the systematic error at 150~MHz as a function of LST is shown in the right panel of Figure~\ref{syser_spect}; the variation is owing to change in antenna temperature with LST.

\subsubsection[Likelihood and Confidence limit]{The goodness of fit for the estimated parameters}
\label{section_likelihood}

We examine the distribution of $\chi^2$ over the 8-dimensional parameter space described in Section~\ref{Modelling},  
denoted by the parameter-space vector $\Theta$,  for evaluating the goodness of fit. 
Given the residual to the fit: the difference between the model temperature 
  ($T_{model}(\nu_{i})$) and the measured temperature 
($T_{a0}^{'}(\nu_{i})$) and 
the measurement error at each frequency ($\sigma(\nu_i)$), the Likelihood function $L(D|\Theta)$ leads to confidence distribution in the parameter space.
Prior ranges for the variation  in parameters were refined in the modeling steps; our final fits and likelihood distributions were found to be insensitive to the actual ranges adopted.  We  marginalize over the six parameters that describe the system to obtain the likelihood contours for the sky parameters $T_0$ and 
$\alpha$. 


\begin{figure*}[ht]
\centering
\includegraphics[angle=0, width=0.5\linewidth]{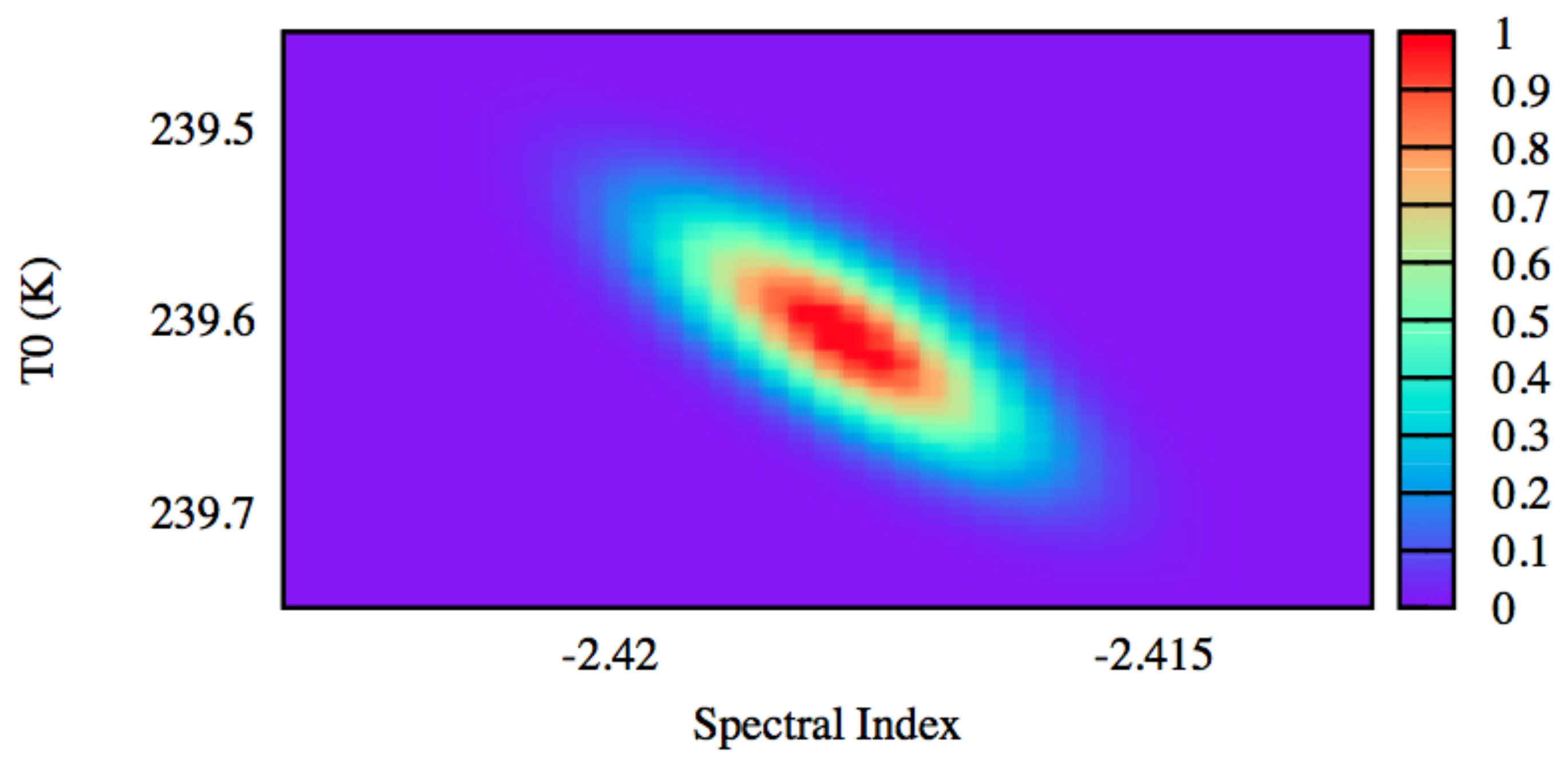}
\caption{Example marginalized likelihood distribution in the parameter space  $\alpha$ versus $T_{0}$ is shown for a 6-min averaged measurement set.
}
\label{Con}       
\end{figure*}

The marginalized confidence region for $T_0$ and $\alpha$ is shown for 
a 6-min averaged measurement set in Figure~\ref{Con}.  The formal 1-standard deviation error in the spectral index, from the goodness of fit, is $\Delta\alpha \simeq 0.01$ and the corresponding error in the absolute sky brightness is a fraction of a K, much smaller than the systematic error in the estimate of this parameter.

\section{Derivation of a normalization for the 150-MHz all sky map}
\label{section_comparison}

\begin{figure*}[ht]
\begin{minipage}[b]{0.5\linewidth}
\centering
\includegraphics[angle=0, width=\linewidth]{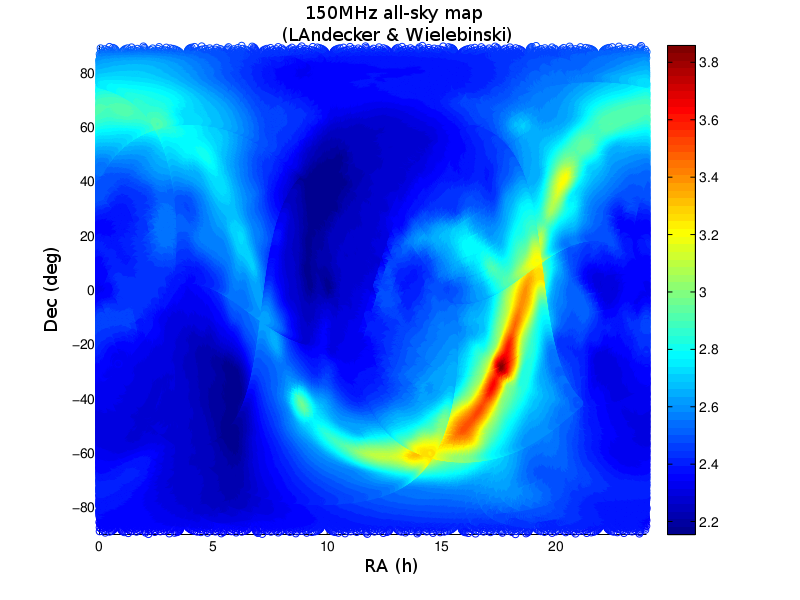}
\end{minipage}
\hspace{0.1cm}
\begin{minipage}[b]{0.5\linewidth}
\centering
\includegraphics[angle=0, width=\linewidth]{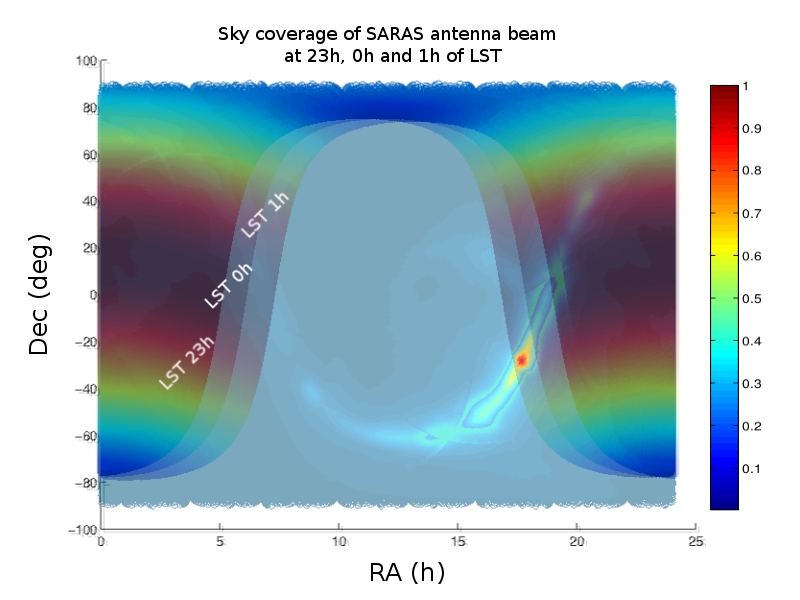}
\end{minipage}
\caption{On the left is shown the 150 MHz all-sky map of \citep{Landecker70}. The log$_{10}$ sky brightness temperature is shown in color-coded intensity. 
On the right is shown the sky regions covered by the SARAS beam at LST 23$^h$, 0$^h$ and 1$^h$ response of the SARAS antenna beam at various RA and dec is overlaid on the sky-map showing the sky coverage. Color palette shows the antenna gain at various angles.}
\label{sky_coverage}       
\end{figure*}

\begin{figure*}[ht]
\begin{minipage}[b]{0.5\linewidth}
\centering
\includegraphics[angle=270, width=0.7\linewidth]{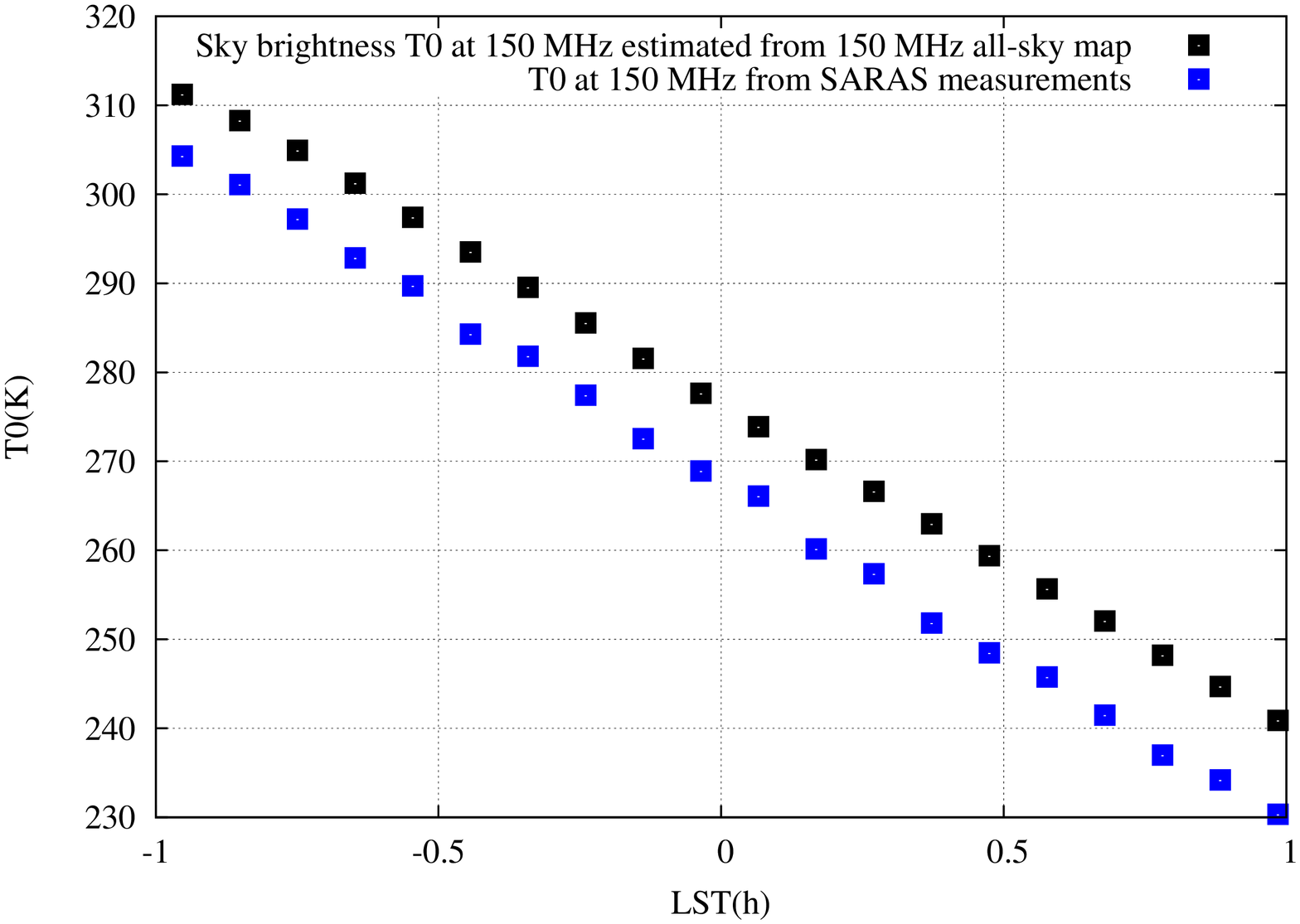}
\end{minipage}
\begin{minipage}[b]{0.5\linewidth}
\centering
\includegraphics[angle=270, width=0.7\linewidth]{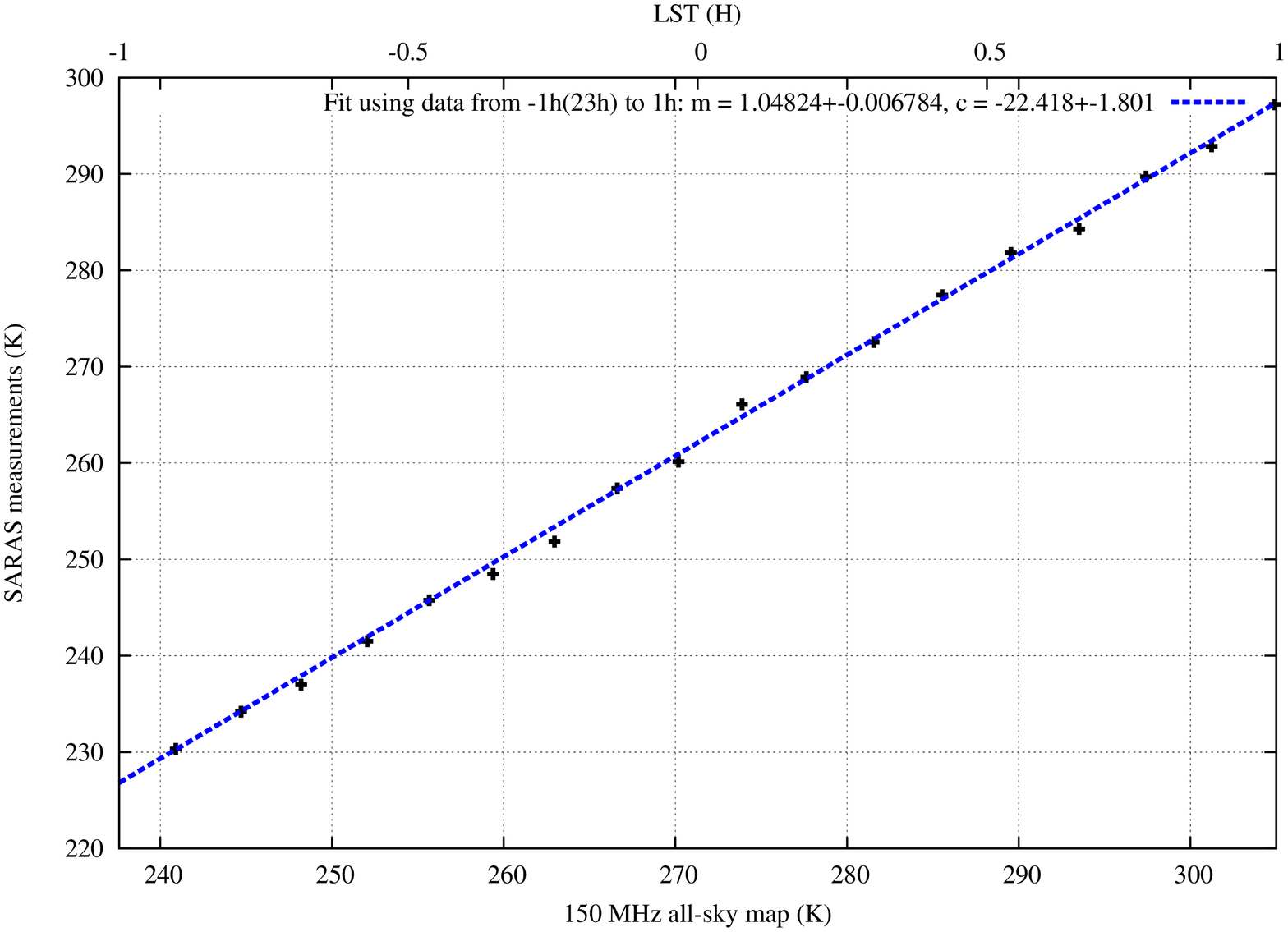}
\end{minipage}
\caption{On the panel on the left is a comparison of the SARAS measurements of sky brightness temperature and expectations for this measurement derived using the 150-MHz all-sky map of \citet{Landecker70}.  In the panel on the right is a temperature-temperature plot of the SARAS measurement versus the corresponding expectation from the Landecker \& Wielebinski map.}
\label{150_comp}       
\end{figure*} 

Corresponding to the observing site and LST range in which the SARAS measurements were done, the expected beam-averaged sky brightness was computed by smoothing the all-sky 150~MHz map \citep{Landecker70} to the frequency independent beam of SARAS, which is that of a short dipole.  In Fig.~\ref{sky_coverage} is shown the sky areas covered by the SARAS beam over the LST range of the observing.
This expectation for the sky temperature variation over LST is shown in Fig.~\ref{150_comp} along with the results from the SARAS measurements.  
SARAS measurements are lower than the expectations from the Landecker \& Wielebinski maps; however, the difference is within the quoted offset and scale
errors of 50~K and 5\% respectively for the 150-MHz map.

The accuracy of absolute calibration of the SARAS system is $0.6\%$. The systematic error in estimating the 150-MHz sky brightness is 6.15~K  at LST 23$^{h}$ and 5.7~K at LST 1$^{h}$.  The half-width of the likelihood distribution for the estimated sky brightness temperature $T_{0}$ varies between $\pm 0.05$ to $\pm 0.2$~K, based on the marginalized likelihoods.   The error in the best-fit value of the spectral index $\alpha$, estimated from the marginalized likelihood function, varies between $\pm 0.001$ and $\pm 0.005$.  Accounting for all of these, the uncertainty in the estimated sky brightness and spectral index within the SARAS observing band is $<1\%$. The rms of the residuals in the real and the imaginary parts of the calibrated measurement set, after the final fit, is 1.45 and 0.72~K ($\approx 0.5\%$).  Table~2 summarizes the expectation for the error in the measurement of the mean sky brightness temperature $T_{0}$ at 150~MHz by SARAS.

\begin{table}[ht]
\caption{Sources of error in SARAS measurement of sky brightness at 150~MHz. }
\vspace{2em}
\label{table:error}
\centering
\begin{tabular}{lr}
\hline\noalign{\smallskip}
Source of error & Magnitude (\%)\\
\noalign{\smallskip}\hline\noalign{\smallskip}
Absolute calibration error & 0.6   \\
Formal error in best-fit parameter $T_0$ & 0.1\\
Systematic error & 1.2  \\
\noalign{\smallskip}\hline
\label{result_tabular}
\end{tabular}
\end{table} 

To compare the SARAS measurements with the 150 MHz all-sky map a temperature-temperature plot was made; this is shown in Fig.~\ref{150_comp}.   A straight-line fit of the form $y = mx+c$ to this plot yields a scaling constant $m=(1.04824\pm0.006784)$ and a zero point offset $c=(-22.418\pm1.801)$~K between the two measurements.  Zero point error in 150 MHz all-sky map is quoted to be $\pm50$~K whereas the estimated error in the temperature scale is $5\%$ \citep{Landecker70}.  If the 150-MHz map of Landecker \& Wielebinski is scaled by factor 1.04824 and its zero point is corrected by subtracting a value of $-22.418$, the resulting map would have improved its absolute calibration.  
 Since absolute calibration accuracy of SARAS is $\>0.6\%$, scaling by a constant $m$ improves the accuracy of the absolute temperature scale of the 150-MHz map to $0.7\%$.  The zero point uncertainty in the SARAS measurement is dominated by the systematic error in the measurement, whose 
median value  is $\pm5.9$~K between 23 and 1$^h$ (Figure \ref{syser_spect}).  Hence the suggested recalibration of the 150~MHz map also reduces the zero point uncertainty  in the map to $\pm8$~K.


 
 In Fig.~\ref{Step3_param_LST} we show the the change in the value of $\alpha$ as LST increases; this change is 
significant and larger than the estimated error on $\alpha$. In other words, 
we report a statistically significant variation of $\alpha$ over the LST range 23 to 01$^h$. 

\section{Discussion and Summary}

SARAS is a correlation spectrometer designed to measure the radio continuum sky brightness at between 87.5 and 175~MHz \citep{Patra13}. In this paper we presented the analysis results of SARAS observations made from the Gauribidanur radio observatory, Bangalore, India, between 23 and 1$^h$ LST. At the observing site where SARAS was deployed frequencies below 110~MHz were affected by FM stations and hence the observations were limited to the 110--170~MHz band. 

Absolute calibration of SARAS was done using a broadband calibration noise source which was calibrated with the noise power from a well matched thermal load of known physical temperature.  We adopt a 8-parameter model, six parameters to describe the additive systematics along with two parameters to describe the sky spectrum and adopt a hierarchical approach to optimize the fit and determine the parameters. 

The measured variation of sky brightness at 150~MHz from SARAS  is compared expectations based on the 150-MHz all-sky map of \citet{Landecker70}, which was smoothed down to the SARAS beam.  The comparison confirms the correctness of the calibration of the map of  \citet{Landecker70} within their quoted errors.  More importantly, the SARAS measurement yields an improved calibration for the map: scaling by factor 1.04824 and correcting its zero point by subtracting a value of $-22.418$ yields a 150-MHz with zero-point error of  8~K and scale error of 0.8\%.

The estimated spectral index shows a slow variation with LST with spectral index steepening from $\alpha = -2.3$ to $-2.45$ between 0 to 1$^h$ LST.  As seen in Fig.~\ref{sky_coverage}, the Galactic Centre region progressively moves out of the dipole beam as time progresses and the LST at the observatory changes from 23 to 0$^h$.  This indicates that the spectral index of the radio sky is steeper off the Galactic Center region and perhaps off the Galactic plane..
This is consistent with measurements  by \cite{Turtle62} who observed a small difference in the spectral index between the direction of the Galactic halo and the disk. Steepening of spectral index towards the galactic halo has also been observed at  higher frequencies  \citep{Bennett03, Fuskeland14, Platania98}. This  observation is consistent with what is expected in a simple diffusion model where high energy cosmic ray electron originate in star-forming activity and consequent supernovae in the Galactic disk and diffuse into the Galactic halo, with spectral steepening owing to synchrotron losses and inverse Compton scattering. 

\textbf{Acknowledgements:}
The SARAS system was deployed at the Gauribidanur Radio Observatory for the observations presented herein; the observations have been enabled by substantial logistics and engineering assistance by the staff of the observatory. We acknowledge the engineering efforts of members of the Radio Astronomy Laboratory and Mechanical and Carpentry workshops of Raman Research Institute in design, fabrication, commissioning, systems engineering, system deployment and maintenance. We would also like to thank Professor Ronald D Ekers (CSIRO-CASS, Australia) for his encouragement.\\

\end{document}